\documentclass[12pt,superscriptaddress,showpacs]{revtex4}

\usepackage{dcolumn}% Align table columns on decimal point
\usepackage{bm}% bold math

\usepackage[T1]{fontenc}
\usepackage[latin1]{inputenc}
\usepackage[english]{babel}
\usepackage{amsmath}
\usepackage{amssymb}
\usepackage{amstext}
\usepackage{braket}
 \usepackage[pdftex]{graphicx}
\usepackage{relsize}
\usepackage{subfigure}
\newcommand{\ud}{\mathrm{d}}
\newcommand{\ue}{\mathrm{e}}

\newcommand{\vs}[1]{\ensuremath{\boldsymbol{#1}}}
\newcommand{\vect}[1]{\ensuremath{\bm{#1}}}

\begin{document}

%\preprint{}

\title[Modified spin-wave theory for frustrated Heisenberg antiferromagnets]
{Modified spin-wave theory with ordering vector optimization II: Spatially anisotropic triangular lattice and $J_1J_2J_3$ model with Heisenberg interactions}
\author{Philipp Hauke}
\address{ICFO -- Institut de Ci\`encies Fot\`oniques, Av.\ Canal Ol\'impic s/n, E-08860 Castelldefels (Barcelona), Spain}
\email{Philipp.Hauke@icfo.es}
\author{Tommaso Roscilde}
\address{Laboratoire de Physique, Ecole Normale Sup\'erieure de Lyon, 46 All\'ee d'Italie, F-69007 Lyon, France}
\author{Valentin Murg}
\author{J.\ Ignacio Cirac}
\address{Max-Planck-Institut f\"ur Quantenoptik, Hans-Kopfermann-Str.\ 1, D-85748 Garching, Germany}
\author{Roman Schmied}
\address{Department of Physics, University of Basel, Switzerland}
%\ead{Philipp.Hauke@icfo.es}

\date{\today}

\begin{abstract}

We study the ground state phases of the $S=1/2$ Heisenberg quantum antiferromagnet on the spatially anisotropic triangular lattice and on the square lattice with up to next-next-nearest neighbor coupling (the $J_1J_2J_3$ model), making use of Takahashi's modified spin-wave (MSW) theory supplemented by ordering vector optimization.
We compare the MSW results with exact diagonalization and projected-entangled-pair-states calculations, demonstrating their qualitative and quantitative reliability. 
We find that MSW theory correctly accounts for strong quantum effects on the ordering vector of the magnetic phases of the models under investigation: in particular collinear magnetic order is promoted at the expenses of non-collinear (spiral) order, and several spiral states which are stable at the classical level, disappear from the quantum phase diagram. 
Moreover, collinear states and non-collinear ones are never connected continuously, but they are separated by parameter regions in which MSW breaks down, signaling the possible appearance of a non-magnetic ground state. In the case of the spatially anisotropic triangular lattice, a large breakdown region appears also for weak couplings between the chains composing the lattice, suggesting the possible occurrence of a large non-magnetic region continuously connected with the spin-liquid state of the uncoupled chains. 
\end{abstract}

%    75.30.Ds Spin waves
%    75.30.Kz Magnetic phase boundaries (including magnetic transitions, metamagnetism, etc.)
%    75.10.Jm Quantized spin models
%    75.50.Ee Antiferromagnetics

\maketitle

%%%%%%%%%%%%%%%%%%%%%%%%%%%%%%%%%%%%%%%%%%%%%%%%%%%%%%%%%%%%%%%%%%%%%%%%%%%%%%%%%%%%%%%%%%%%%%%%%%%%%%%%%%%%%%%%%%%%%%%%

\section{Introduction}

Low-dimensional frustrated quantum spin systems can display an intriguing interplay between order and disorder: classical order has been shown to be quite resilient in two or three dimensions \cite{Dyson1978,Kennedy1988,Manousakis1991,Misguich2004}; frustration, however, can lead to the melting of magnetic long-range order (LRO) and the emergence of quantum disordered states like valence-bond solids or resonating valence bond 
states \cite{Anderson1973,Fazekas1974}. 
Understanding such magnetically disordered quantum phases is important for the search for fractionalized excitations in two dimensions \cite{Anderson1973}, 
as well as for the understanding of the behavior of layered magnetic insulators/metals in which
magnetism is disrupted by charge doping, leading to dramatic phenomena such as
superconductivity at high critical temperature \cite{Kastner1998,Lee2006,delaCruz2008}.

A large variety of magnetic materials can be described by the Heisenberg Hamiltonian
\begin{equation}
 \label{HS}
  H_{\text{S}}=
  \sum_{\braket{i,j}} J_{ij} ~{\bm S}_i\cdot{\bm S}_j,
\end{equation}
where ${\bm S}_i$ is a quantum spin-$S$ operator at site $i$. 
In this paper, we will focus on the antiferromagnetic case for $S=1/2$, and 
on two-dimensional frustrated lattices. Quasi-two-dimensional frustrated antiferromagnetism
is relevant to a variety of $S=1/2$ compounds, realizing the spatially
anisotropic triangular lattice (\emph{e.g.}, in $\mathrm{Cs}_2\mathrm{CuCl}_4$
\cite{Coldea2001} and $\kappa$-(BEDT-TTF)$_2$Cu$_2$(CN)$_3$ \cite{Shimizu2003, Yamashita2008}, etc.), or
the frustrated ($J_1J_2$) square lattice 
(\emph{e.g.}, in Li$_2$VOSi(Ge)O$_4$, VOMoO$_4$ \cite{Carretta2004}, BaCdVO(PO$_4$)$_2$ \cite{Nath2008}, etc.). For both lattice geometries, the Heisenberg model is expected to display spin-liquid phases for particular values of the frustrated couplings, 
although the extent and nature of these spin-liquid phases is still under theoretical debate,
both for the spatially anisotropic triangular lattice (SATL) \cite{Weihong1999,Yunoki2006,Weng2006,
Fjaerestad2007,Kohno2007,Starykh2007,Heidarian2009}
and for the frustrated square lattice 
\cite{Singh1999,Capriotti2001,Sushkov2001,Sindzingre2004,Sirker2006,Mambrini2006,Darradi2008}.

In this work, we investigate the $S=1/2$ Heisenberg antiferromagnetic Hamiltonian on two-dimensional frustrated lattices making use of Takahashi's modified spin-wave (MSW) theory \cite{Takahashi1989}, supplemented with  the optimization of the ordering vector \cite{Xu1991}. In a previous paper \cite{Hauke2010}, we have shown  that (for the SATL with XY interactions) this approach provides a significant improvement over conventional spin-wave theory (as well as over conventional MSW theory), as it allows to correctly account for the dramatic quantum effects occurring to the form of order which appears in frustrated quantum antiferromagnets, and for the quantum corrections to the stiffness of the ordered phase. In particular, a very low stiffness, or the complete breakdown of the theory, provide strong signals that the true ground state might be quantum disordered; hence, this method serves as a viable approach to finding candidate models potentially displaying spin-liquid behavior. 
For a more detailed description of the formalism we refer the reader to Ref.\ \cite{Hauke2010}.
 
Here, we apply this MSW theory with ordering vector optimization to the Heisenberg SATL, as well as to the square lattice with nearest, next-to-nearest and next-to-next-to-nearest neighbor couplings (the $J_1J_2J_3$ model \cite{Figueirido1989,Read1991,Ferrer1993,Mambrini2006}).
Both models feature a very complex  $T=0$ phase diagram, with spirally and collinearly ordered regions, whose ordering vector is subject to strong quantum corrections with respect to the classical ($S\to\infty$) limit.
They also feature extended breakdown regions for MSW theory, pointing at the possible 
 spin-liquid nature of the true ground state of the system.  Comparison with numerical results 
 coming from exact diagonalization and projected-entangled-pair-state (PEPS)
 calculations show that MSW theory correctly accounts for some of the most salient
 features of the quantum phase diagram of these systems, and that it hence represents
 a very versatile tool to probe the robustness (or the breakdown) of a semi-classical 
 description of the ground state of frustrated quantum magnets.  
 %%\comment{Partially repeats previous paragraph.}

The remainder of this paper is organized as follows: 
Section \ref{cha:msw_triangNN_Heis} presents the ground state phase diagram of the SATL 
with nearest-neighbor Heisenberg interactions; in Section \ref{cha:J1J2J3}, we calculate the ground state phase 
diagram of the $J_1J_2J_3$ model; finally, in Section \ref{cha:conclusion} we present our conclusions. 
The technical aspects of MSW theory applied to Heisenberg antiferromagnets are presented in the Appendix.

%%%%%%%%%%%%%%%%%%%%%%%%%%%%%%%%%%%%%%%%%%%%%%%%

\section{\label{cha:msw_triangNN_Heis}MSW theory on the spatially anisotropic triangular lattice with nearest-neighbor Heisenberg-bonds}

\begin{figure}
        \centering
        \includegraphics[width=0.75\textwidth]{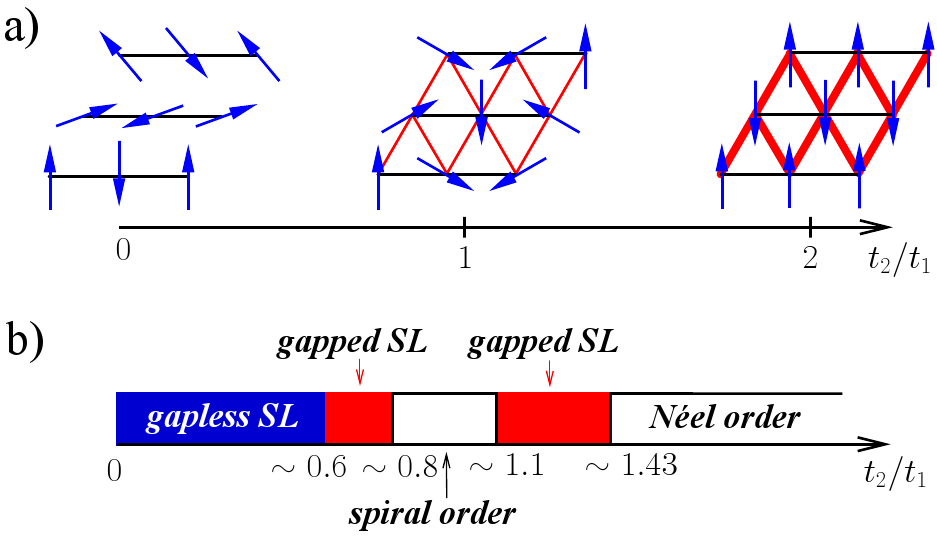}
        \caption{\label{fig:phasediagtriang}
        (a) Classical ground state phase diagram of the SATL with a sketch of the 1D state at $\alpha=0$, the spiral state at $\alpha=1$, and the 2D-N\'{e}el state for $\alpha\geq 2$. (The horizontal black bonds have strength $t_1$, and the diagonal red bonds have strength $t_2$.)
        (b) The quantum mechanical phase diagram changes considerably due to order-by-disorder effects and the appearance of spin liquids \cite{Weihong1999,Yunoki2006}.}
\end{figure}

The triangular lattice with Heisenberg interactions has been considered as one of the first candidate systems for quantum-disordered behavior in the ground state \cite{Anderson1973}.
Recently, the phase diagram of the spatially anisotropic triangular lattice (SATL) up to values of $\alpha\equiv t_2/t_1=1$ has been studied by Yunoki and Sorella using variational quantum Monte Carlo methods \cite{Yunoki2006}. They find that the gapless spin-liquid phase of the isolated chains ($t_2 = 0$) persists also at finite coupling up to a critical value $\alpha\approx 0.65$, followed by a gapped spin liquid; for $\alpha\approx 0.8$ the gap closes and the system undergoes an ordering transition to spiral order, continuously connected with the 3-sublattice order of the isotropic Heisenberg antiferromagnet
($\alpha=1$). 
This scenario is still controversial, however: studies based on low-energy effective field theory
for the description of the coupled chains in the case $\alpha < 1$ indicate that the system might still exhibit long-range antiferromagnetic order even for very weak coupling among the chains. This form of order results from high-order perturbation theory in the inter-chain coupling, and it is necessarily very weak, given that numerical methods cannot detect it. Its observation is clearly beyond the capabilities of our MSW approach.
Coming from the large-$\alpha$ limit, series expansions by Weihong \emph{et al.}\ indicate that 2D-N\'{e}el order  -- appearing on the square lattice defined by the dominant $t_2$-couplings -- persists down to $\alpha\simeq 1.43$, followed by a phase without magnetic order in the interval $1.1\lesssim\alpha\lesssim 1.43$ \cite{Weihong1999}. Below this region the authors find incommensurate spiral order connecting continuously to the isotropic point $\alpha =1$. 
In Ref.\ \cite{Manuel1999}, qualitative similar results have been obtained using the Schwinger-boson approach. 
The resulting phase diagram differs strongly from the classical one, which is characterized by spiral order for $0 < \alpha < 2$, and by N\'eel order
for $\alpha \geq 2$.  
The classical phase diagram  is contrasted with the quantum mechanical one (composed from Refs.~\cite{Weihong1999} and~\cite{Yunoki2006}) 
in Fig.~\ref{fig:phasediagtriang}. It is interesting to notice that a qualitatively similar phase diagram has been obtained recently by some of us for 
the XY model on the SATL \cite{Schmied2008, Hauke2010}. 

A variety of experiments have been carried out on magnetic compounds described by the Heisenberg model on the SATL, with results that are still controversial.
For instance neutron scattering experiments of Coldea and coworkers \cite{Coldea2001} on $\mathrm{Cs}_2\mathrm{Cu}\mathrm{Cl}_4$, where $\alpha\approx 1/3$, claimed evidence that the low-energy physics is governed by spinons, fractionalized excitations with $S=1/2$ which represent the elementary excitations
in the case of uncoupled chains. Yet, Ref.~\cite{Kohno2007} showed that, for a finite inter-chain coupling, spinons tunnel between chains in bound pairs with 
$S=1$ (so-called triplons), so that the fractionalization in two dimensions is strictly speaking not present. 
Ref.~\cite{Kohno2007} argues that the spinons in $\mathrm{Cs}_2\mathrm{Cu}\mathrm{Cl}_4$ are descendants of the excitations of the individual 1D chains and not characteristic of any exotic 2D state. This further reinforces the idea of a quasi one-dimensional behavior up to relatively high inter-chain interactions mentioned in the previous paragraph.

\subsection{MSW predictions for the ground-state phase diagram}

In this section, we discuss the ground-state phase diagram resulting from the predictions of MSW theory for the $S=1/2$ SATL with nearest-neighbor (NN) Heisenberg interactions.

\begin{figure}
        \centering
        \includegraphics[width=0.25\textwidth]{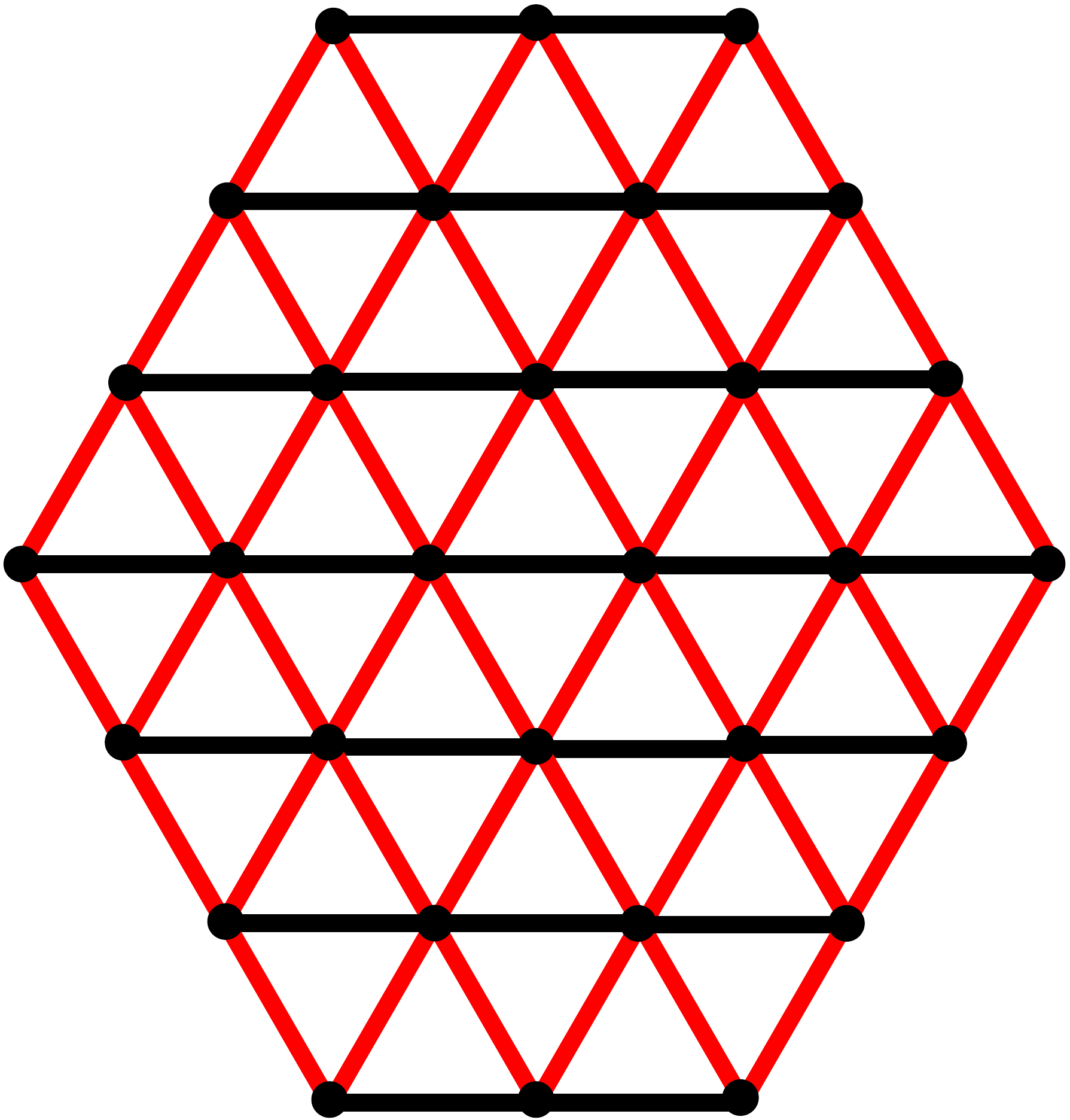}
        \caption{Cluster of 30 spins for which we carried out ED. The 24-spin system is equivalent, only with the top and bottom rows removed. Black dots denote sites, the horizontal black bonds have strength $t_1$, and the diagonal red bonds have strength $t_2$. \label{fig:systemsexactdiag}
        }
    \end{figure}

In order to assess the validity of MSW results, we compare them with exact diagonalizations (ED). Using the Lanczos method, we compute the ground state of small clusters of 14, 24, and 30 spins. The considered geometry for the 30-spin system can be found in Fig.~\ref{fig:systemsexactdiag}. The 24-spin system can be obtained from it by removing the top and bottom rows. The 14-spin cluster is an equivalent system with rows of 2, 3, 4, 3, and 2 spins. The clusters are chosen for their symmetry with respect to reflection along the coordinate axis, and for their ratio of the number of $t_2$-bonds (red) to the number of $t_1$-bonds (black), which lies
close to the bulk value of 2. We use open boundary conditions to allow for the accomodation of spiral order with arbitrary wave vector. 
   
We find that, due to the peculiar geometries chosen, there exist parameter ranges where the ground state falls into the threefold degenerate triplet with total spin equal to unity. Nonetheless, we restrict our calculations to the $M_z^{\mathrm{total}}=0$ subspace (with $M_z^{\mathrm{total}}$ being the $z$ component of the total spin), and the $M_z^{\mathrm{total}}=\pm 1$ states are excluded. This results in an apparent breaking of the $x$--$z$ symmetry (the $x$--$y$ symmetry is preserved). This symmetry would be recovered by averaging over the whole triplet subspace. The reason for such an apparent symmetry breaking resides in the particular geometry of the cluster considered, which complicates the comparison between different system sizes. This triplet physics might play an important role for bigger systems, although one cannot draw conclusions about the thermodynamic limit from the small clusters considered. A non-trivial triplet physics could be especially an issue for variational methods restricting their focus to the singlet subspace. 

The lattice sizes considered in the MSW calculations are $32\times 32$ spins and the infinite lattice limit, which is achieved by transforming sums over the first Brillouin zone into integrals. Figures~\ref{fig:E0_Heis_triang} to~\ref{fig:CC_Heis_triang} show that the difference between the lattice sizes is insignificant except near quantum phase transitions, which is expected because of the divergence of correlation lengths near criticality.

\subsubsection{Parameter regions where MSW theory fails to converge.} 
Convergence in the self-consistent equations of MSW theory with ordering vector optimization, 
  Eqs.~(\ref{tanh2th}--\ref{constr2},~\ref{Qx2}), 
  cannot be achieved in selected regions of the ground state phase diagram, namely
  for $\alpha \lesssim 0.65$ and for $1.14 \lesssim \alpha \lesssim 1.3$. 
  (Interestingly, convergence is restored in the pure 1D limit, $\alpha=0$, 
  for which the theory formulates surprisingly good predictions.)  
  This breakdown of convergence corresponds to the appearance of an imaginary part in the spin-wave frequencies, Eq.~\eqref{disp}, signaling an instability of the ordered ground state.  
  The breakdown of a self-consistent description of the system in terms of an ordered ground state is strongly suggestive
  of the presence of a quantum-disordered ground state in the exact behavior of the system. 
  Hence, one can interpret these parameter regions as candidates for the spin-liquids predicted 
  from Refs.~\cite{Weihong1999,Yunoki2006} [compare Fig.~\ref{fig:phasediagtriang}~(b)]. 
 Both for $\alpha < 1$ and $\alpha > 1$, we find that the breakdown region of MSW appears to be fully contained within the 
 region of SL behavior (either gapped or gapless) estimated in Refs.~\cite{Weihong1999,Yunoki2006}. 
 Hence MSW theory is seen to possibly underestimate the width of the quantum-disordered
 regions in the phase diagram, which is to be expected due to the partial account of quantum
 fluctuations given by MSW theory.

\subsubsection{MSW ground state energy in comparison with previous results.}
\begin{table*}
{\scriptsize
\begin{tabular}{|l||l*{4}{|l}||l|}
  \hline
  Method  \qquad  & $\alpha=0$ & $\alpha=0.7$ & $\alpha=0.8$ & $\alpha=1$ & $\alpha=\infty$ & $M$ at $\alpha=1$\\
  \hline
  \hline
  exact, thermodynamic limit                    & $-0.443147$   &               &               &               &  					&  							\\
  exact, $N=30$ (present study)                 & $-0.4127$   	&               &               & $-0.5471$     &  					& $0.1314$ 			\\
  exact, $N=40$, extrapolated \cite{Richter2010} & 					    &               &               & 					    & $-0.6701$ & 							\\
  VMC (RVB) \cite{Yunoki2006}                    &               &               &               & $-0.5123(1)$ 	&  					& $0.0$ 				\\
  VMC (RVB with $\mu=0$) \cite{Yunoki2006}       &               &               &               & $-0.5291(1)$  & 					& $0.0$ 				\\
  VMC (BCS+spiral) \cite{Weber2006}            	&               &               &               & $-0.532(1)$   &  					& $0.36$ 				\\
  VMC (p-BCS) \cite{Yunoki2006}                 	& $-0.442991$   & $-0.46467$    & $-0.47840$    & $-0.5357(1)$  &  					& $0.0$ 				\\
  FN \cite{Yunoki2006}                           &               & $-0.47051$    & $-0.48521$    & $-0.53989(3)$ &  					& $0.1625(30)$ 	\\
  FNE \cite{Yunoki2006}                          &               &$-0.47171$     & $-0.48691$    & $-0.54187(6)$	&  					& $0.1765(35)$ 	\\
  GFMCSR \cite{Singh1989a,Capriotti1999a}    		&               &               &               & $-0.545(2)$   &  					& $0.205(10)$ 	\\
  series expansion \cite{Singh1989a} \footnotemark[1]              &               &               &               &      & $-0.6696(3)$   & \\
  LSW \cite{Yunoki2006,Singh1989a}\footnotemark[1] 			    	&               &               &               & $-0.538(2)$   &  					& $0.2387$			\\
  MSW (present study)\footnotemark[1]         	& $-0.4647$     & $-0.4639$     & $-0.4775$     & $-0.5303$     & $-0.6699$ & $0.3426$			\\
  \hline
\end{tabular}}
{\scriptsize{\footnotemark[1] These methods do not provide a rigorous upper bound for the ground state energy.}}

\caption[Comparison of the ground state energy of various methods]{\label{tab:E0_Heis_triang_variousmethods}
Comparison of the ground state energy per spin derived by various methods, for some values of $\alpha$. VMC stands for `variational quantum Monte Carlo' where the wave function used is given in brackets \cite{Yunoki2006,Weber2006}. FN is short for lattice fixed node and FNE for the improved FN effective Hamiltonian method \cite{Yunoki2006}.
Furthermore included are the exact result of the Heisenberg chain in the thermodynamic limit, the ED results for the 30-spin cluster, and the ED results from finite size extrapolations of calculations on clusters of up to 40 spins \cite{Richter2010}.
Also given are the estimates of LSW theory from Ref.~\cite{Yunoki2006} and the Green's function Monte Carlo method with stochastic reconfiguration (GFMCSR) \cite{Capriotti1999a}. The last column gives the staggered magnetization or, in the case of MSW theory, the population of the zero mode $M_0$.}
\end{table*}
\begin{figure}
        \centering
        \includegraphics[width=0.99\textwidth]{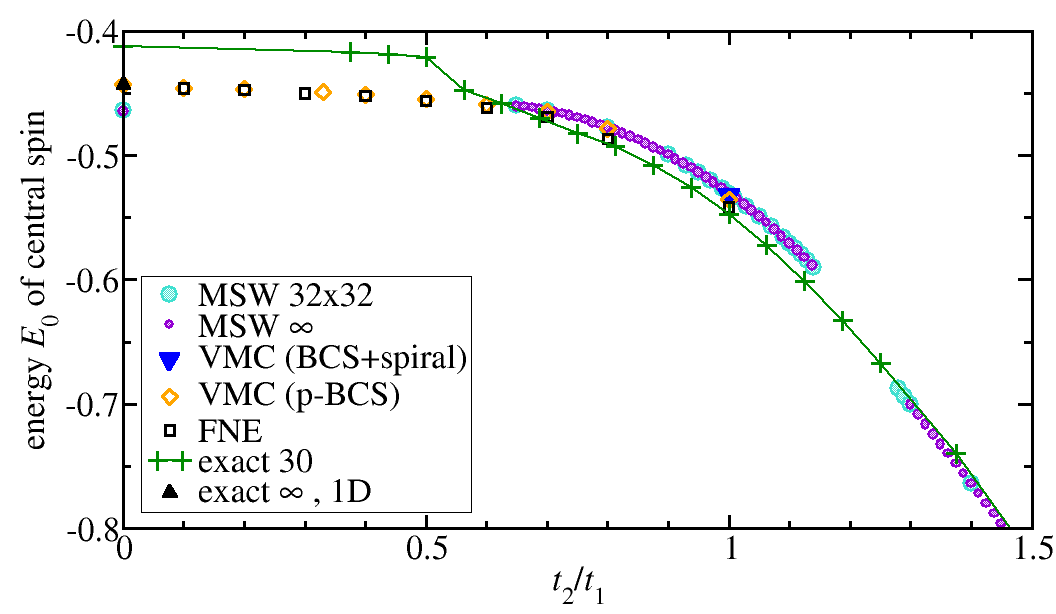}
        \caption{        \label{fig:E0_Heis_triang}
        MSW results for the ground state energy lie close to results from previous studies. Shown are the data of Ref.~\cite{Yunoki2006} for their variational quantum Monte Carlo (VMC) \emph{Ansatz} with a projected BCS wave-function (p-BCS) and the improved FN effective Hamiltonian method (FNE). We further display the value obtained in the isotropic limit by Ref.~\cite{Weber2006} by use of a VMC method with a mixture of a BCS and a spiral ordered wave-function (BCS+spiral), and the exact result of the 1D limit. The numbers in the labels of the curves are the respective system sizes considered.
        }
\end{figure}
Table~\ref{tab:E0_Heis_triang_variousmethods} demonstrates that the energy from MSW theory compares very well to results that were obtained recently by Yunoki and Sorella by variational quantum Monte Carlo methods \cite{Yunoki2006}, also plotted in Fig.~\ref{fig:E0_Heis_triang}. For comparison, we also show the curve that they obtain with a projected-BCS (p-BCS) wave-function.
In the isotropic triangular lattice, the MSW energy compares also favorably to the data from the Green's function Monte Carlo method with stochastic reconfiguration (GFMCSR) from Ref.~\cite{Capriotti1999a}, but both energy and order parameter (see section~\ref{cha:msw_triangNN_Heis_M0gc}) lie closest to the variational quantum Monte Carlo calculation from Weber \emph{et al.} ~\cite{Weber2006}, who used a mixture of a BCS wave-function and a wave function with spiral order as their starting point (BCS+spiral). 

At $\alpha=0$ the MSW value $E_0\left(\alpha=0\right)=-0.4647$ is relatively close to the exact result of the one-dimensional case, $-(\ln 2-1/4)=-0.44315$. However, it is located \emph{below} the exact value. This apparent puzzle is easily resolved by noticing that the MSW method is not variational due to the incomplete inclusion of the kinematic constraint (see Appendix). 
We also notice that the ground state energies derived from ED of the 30-site system lie very close to the values from the other methods except in the 1D phase. This could be attributed to the small system size: 
if the interpretation is correct that for small $\alpha$ the Heisenberg SATL is in a 1D-like phase with algebraic correlations, it is natural that finite size effects play a very important role in the critical 1D phase. This would explain the strong deviation of the ED energy in that parameter region. 

On the square lattice ($\alpha\to\infty$), Takahashi showed already twenty years ago the extremely good performance of MSW theory \cite{Takahashi1989}: its ground state energy per spin is $-0.6699$, which is in excellent agreement with the QMC result $-0.669437(5)$ \cite{Sandvik1997}.  

\subsubsection{\label{cha:msw_triangNN_Heis_M0gc}Order parameter and spin stiffness from MSW theory.}

Our next step is to determine the regions where the presence of a finite order parameter $M_0$ and spin stiffness $\Upsilon$ reveal 
magnetic long-range order (LRO). 
Even when $M_0$ and $\Upsilon$ are finite, a caveat is still in order:
a finite order parameter with a very small stiffness might suggest that taking quantum fluctuations more completely into account 
than in MSW theory could lead to a completely disordered state. 

\begin{figure}
        \centering
        \includegraphics[width=0.75\textwidth]{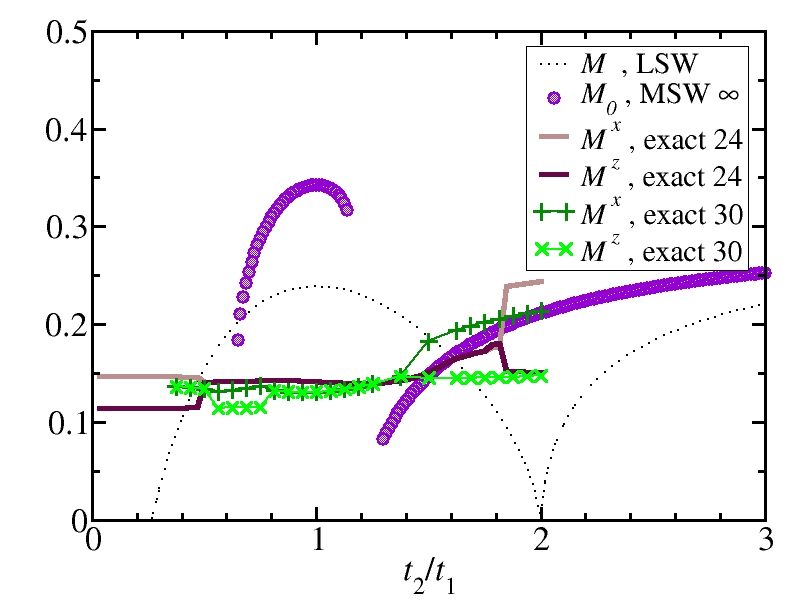}
        \caption[Order parameter]{$M_0$ from MSW theory compared with the LSW values and ED results (see section~\ref{cha:ED}). The numbers in the labels of the curves are the respective system sizes considered in the calculations.}
        \label{fig:M_Heis_triang}
    \end{figure}

\begin{figure}
        \centering
        \includegraphics[width=0.75\textwidth]{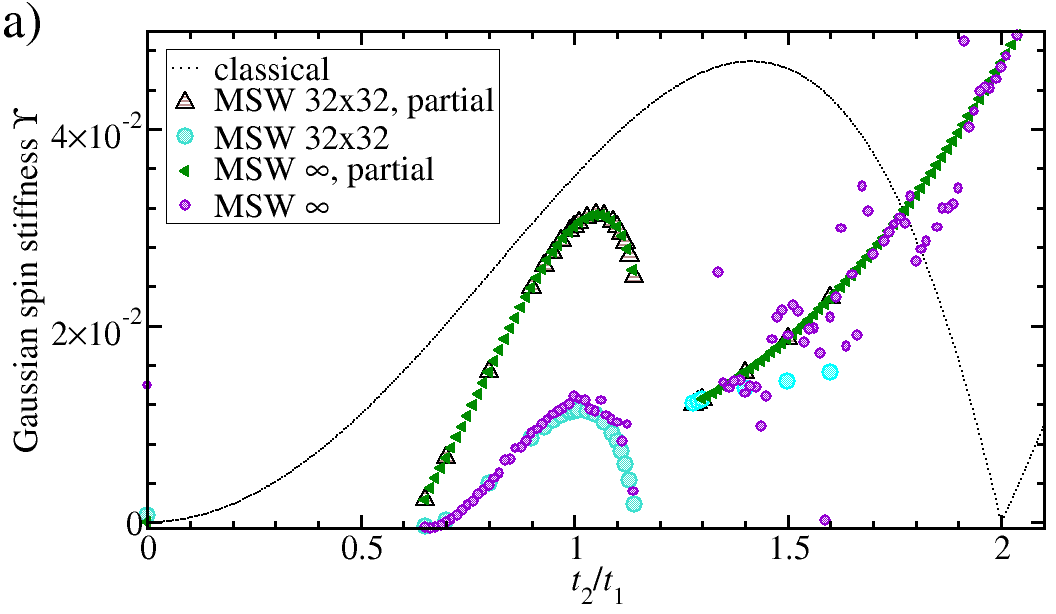}\\
        \includegraphics[width=0.75\textwidth]{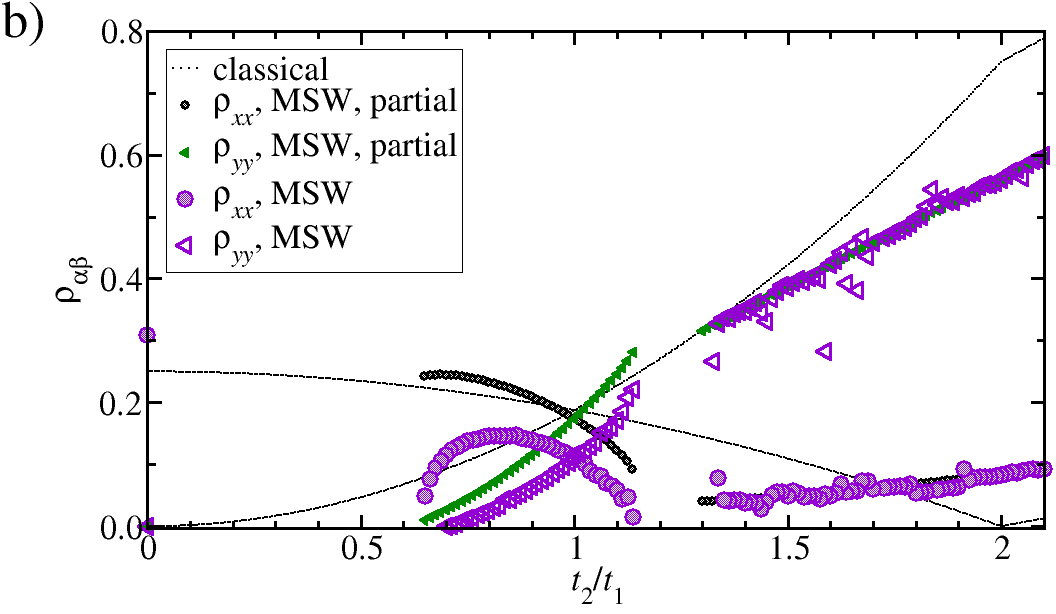}
        \caption[Gaussian spin stiffness]{(a) Gaussian spin stiffness $\Upsilon$ (for the $32\times 32$ and the infinite system) and (b) the components of the spin stiffness tensor (for the infinite system). The mixed second derivative $\rho_{xy}$ vanishes for symmetry reasons. The curves labeled `partial' were obtained by application of Eq.~\eqref{gcpartial}.}
        \label{fig:gc_Heis_triang}
\end{figure}

The order parameter $M_0$, drawn in Fig.~\ref{fig:M_Heis_triang}, shows that magnetic LRO is present
in the intervals $0.65 < \alpha < 1.14$ and $\alpha > 1.3$. This is to be contrasted with linear SW (LSW) theory, 
which predicts the breakdown of magnetic order only for $\alpha \lesssim 0.3$ \cite{Trumper1999}.   
However, in the isotropic case, $\alpha =1$, MSW theory predicts a stronger order parameter than what is predicted
by LSW, as well as by most of the other numerical estimates, which are presented in Table \ref{tab:E0_Heis_triang_variousmethods}.
In the square lattice limit, $\alpha\to\infty$, on the other hand, both MSW and LSW theory attain a staggered magnetization of $0.303$, which compares favorably with the most recent estimates  $M_0=0.311$, based upon diagonalizations of small  clusters of up to 40 spins \cite{Richter2010}.
The MSW order parameter drops drastically upon approaching the region $1.14\lesssim\alpha\lesssim 1.3$ and when reaching the region $\alpha\lesssim 0.65$ from above, the regions where the self-consistent description breaks down, further corroborating the assumption that in these regions magnetic LRO disappears in the true quantum ground state. 
This assumption is strongly reinforced by considering the Gaussian spin stiffness (Fig.~\ref{fig:gc_Heis_triang}): It vanishes at $\alpha=0.65$ and it drops significantly 
when approaching $\alpha=1.14$ from below.

There are various special cases of the SATL for which the spin stiffness has been calculated previously.
In the square lattice limit, $\alpha \to \infty$, MSW theory gives $\rho_{\|}/\alpha=0.216$, somewhat overestimating
the value from QMC $\rho_{\|}/\alpha=0.175(2)$ \cite{Sandvik1997}.
In the isotropic triangular lattice, $\alpha=1$, the spin stiffness from the MSW approach is $\rho_{\|}/\alpha=0.113$.
This value falls between the LSW spin stiffness $\rho_{\|}/\alpha=0.122$ (Ref.~\cite{Lecheminant1995}) and the estimate obtained 
from ED calculations after finite size extrapolation, $\rho_{\|}/\alpha=0.075$ \cite{Lecheminant1995}.
In the limit of decoupled chains, $\alpha = 0$, MSW theory achieves convergence (which was
lost in the interval $0 < \alpha < 0.65$) and provides a spin stiffness $\rho_{xx}/\alpha= 0.309$ in the thermodynamic limit,
relatively close to the exact result in the thermodynamic limit, $\rho_{xx}/\alpha=1/4$ \cite{Shastry1990}.

\subsubsection{Spin and chirality correlations from MSW theory}

\begin{figure}
        \centering
        \includegraphics[width=0.75\textwidth]{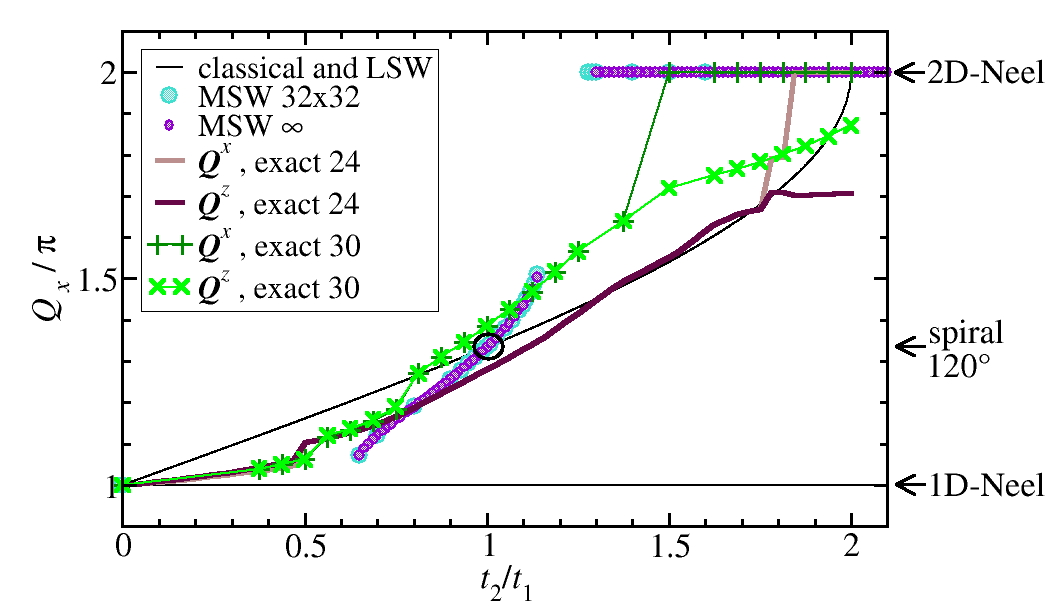}
        \caption{        \label{fig:Q_Heis_triang}
        First component of the ordering wave-vector, $Q_x$, from MSW theory. Also shown are the classical values and the ED results. The black circle marks the order vector $Q_x=120^{\circ}$ of the isotropic triangular lattice which is attained classically and by the spin-wave theories at $\alpha=1$. The numbers in the labels give the system sizes.
        }
\end{figure}

\begin{figure}
				\centering
        \includegraphics[width=0.75\textwidth]{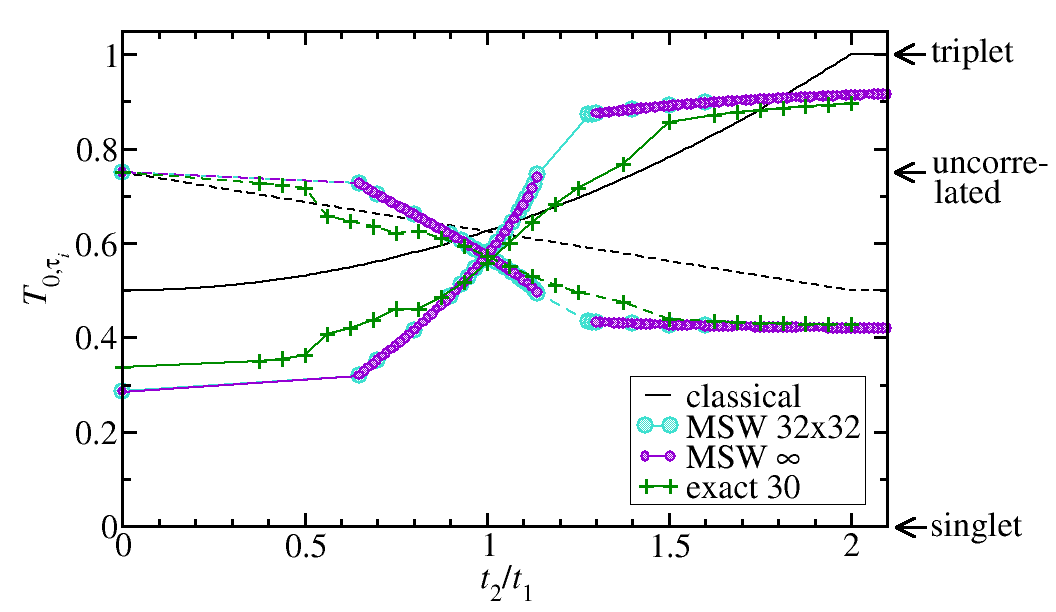}
        \caption{        \label{fig:SS_Heis_triang}
        MSW and ED results show similar behavior of $T_{0,\vect{\tau}_i}$, with $\vect{\tau}_i=\vect{\tau}_1\equiv\left(1,0\right)$ (solid lines) and $\vect{\tau}_i=\vect{\tau}_2\equiv\left(1/2,\sqrt{3}/2\right)$ (dashed lines), respectively. The numbers in the labels give the system sizes.
        }
\end{figure}

\begin{figure}
        \centering
        \includegraphics[width=0.75\textwidth]{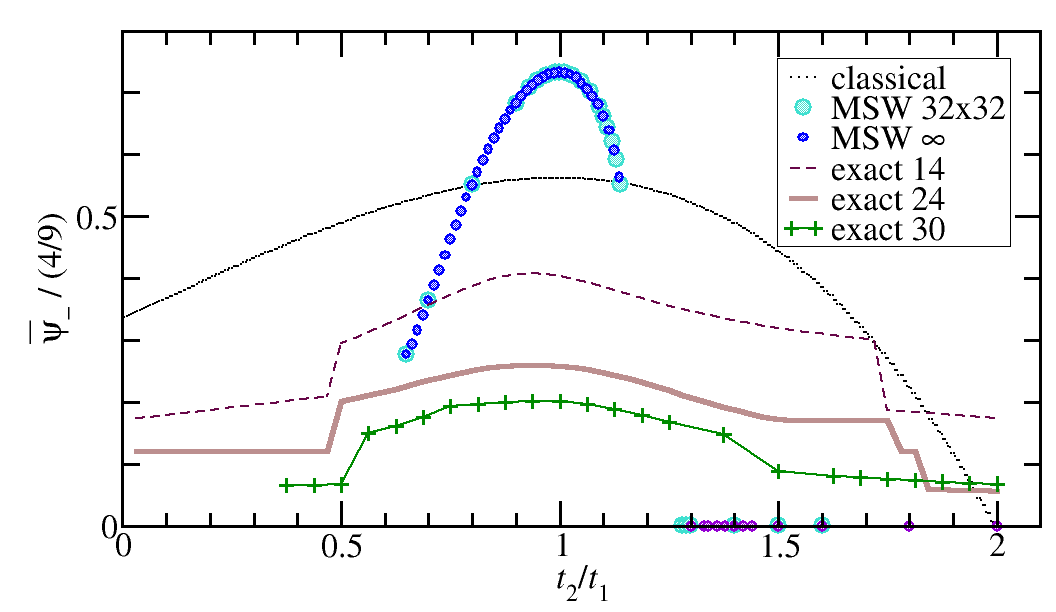}
        \caption{        \label{fig:CC_Heis_triang}
        Comparison of the MSW and ED results for the mean chiral correlation normalized to the theoretical maximum of $4/9$. The numbers in the labels give the system sizes.
        }
\end{figure}

Now we describe the ordered phases found by the MSW \emph{Ansatz} for the Heisenberg SATL in more detail. 
To this end, we analyse the following quantities 
\begin{enumerate}
\item The ordering vector $\vect{Q}$ (Fig.~\ref{fig:Q_Heis_triang}).
Three limiting values for the ordering vector are known. For $\alpha=0$ intra-chain antiferromagnetic (N\'{e}el) order is described by $\vect{Q}=\pi\hat x$. For $\alpha\to\infty$ square-lattice N\'{e}el order is described by $\vect{Q}=2\pi\hat x$. In the isotropic lattice ($\alpha=1$), the threefold symmetry forces the ordering vector to $\vect{Q}=\frac{4 \pi}{3}\hat x$;

\item The spin--spin correlations (Fig.~\ref{fig:SS_Heis_triang}).
We analyze the spin--spin correlations of nearest neighbors through the two-site total spin,
\begin{equation}
\label{totalspinTij}
T_{ij}\equiv\frac 1 2 \braket{\left(\vect{S}_i + \vect{S}_{j}\right)^2}=\braket{\vect{S}_i \cdot \vect{S}_{j}}+\frac 3 4\,.
\end{equation}
This quantity vanishes if the spins are in a singlet, which is equivalent to perfect anticorrelation, takes the value $\frac 3 4$ if they are uncorrelated, and the value $1$ if the spins form a triplet, which means perfect correlation;

\item The mean chiral correlations (Fig.~\ref{fig:CC_Heis_triang}).
Spiral phases carry not only a magnetic order parameter, but also a chiral order parameter. In particular, a vector chirality can be defined on an upwards pointing triangle with counter-clockwise labeled corners $\left(i,j,k\right)$ as \cite{Kawamura2002}
$
\kappa_\Delta=\frac 2 {3\sqrt{3}}\left[\vect{S}_i\times\vect{S}_j+\vect{S}_j\times\vect{S}_{k}+\vect{S}_{k}\times\vect{S}_i\right]_z,
$
and on a downwards pointing triangle with counter-clockwise labeled corners $\left(i,l,j\right)$ as
$\kappa_\nabla=\frac 2 {3\sqrt{3}}\left[\vect{S}_i\times\vect{S}_l+\vect{S}_l\times\vect{S}_{j}+\vect{S}_{j}\times\vect{S}_i\right]_z
$. 
Chirality correlations are defined as \cite{Richter1991}
\begin{equation}
    \label{psi-}
    \psi_-=\braket{\left(\kappa_\Delta-\kappa_\nabla\right)\left(\kappa_{\Delta '}-\kappa_{\nabla '}\right)}\,,
\end{equation}
where the triangle pairs $\left(\Delta,\nabla\right)$ and $\left(\Delta ',\nabla '\right)$ share a $\vect{\tau}_1\equiv\left(1,0\right)$ edge.
In Fig.~\ref{fig:CC_Heis_triang}, we plot the 
average chirality correlation of the central plaquette with all other plaquettes, normalized to the theoretical maximum $4/9$. The MSW data have been obtained by expanding the chiral correlation up to the fourth order in the boson operators, 
which is consistent with the truncation of the bosonic Hamiltonian Eq.~\eqref{H4} to the same order. Going to higher orders does not change the outcome in the regions where $M_0$ is large, but can yield different results where $M_0$ is small.
\end{enumerate}

A comparison of these quantities shows a spiral phase at around $0.65\lesssim \alpha \lesssim 1.14$ and a 2D-N\'{e}el ordered phase for $\alpha\gtrsim 1.3$.
Moreover, when approaching $\alpha\approx 0.65$ from above, the ordering vector, the spin--spin correlations and the ground state energy approach their respective 1D values. This is an indication that below $\alpha\approx 0.65$ the true ground state of the system may enter a 1D-like spin-liquid phase. Nonetheless, 
the vanishing of the spin stiffness for $\alpha \to 0.65^{+}$ is not consistent with the onset of a \emph{gapless} 1D spin-liquid phase, for which the 
spin stiffness should remain finite. Hence, the MSW results rather suggest that the phase appearing below $\alpha = 0.65$
is a \emph{gapped} spin liquid, and that the gapless 1D spin-liquid phase, connected continuously with the limit $\alpha = 0$, is only
attained for even smaller $\alpha$. This seems consistent with the prediction of Ref.~\cite{Yunoki2006} that a gapped
spin-liquid phase separates the spirally ordered phase from the 1D-like gapless disordered one. 

\subsubsection{Order parameter and correlations in comparison with exact diagonalization.\label{cha:ED}}

In the case of ED, the static structure factor
\begin{equation}
    \label{Sk}
    S^\alpha\left(\vect{k}\right)= \frac{1}{N^2} \sum_{i,j} \braket{S_{i}^\alpha S_{j}^\alpha} e^{-i \vect{k} \cdot 	\vect{r}_{ij}}~~~~~~ (\alpha=x,y,z)\,
\end{equation}
allows to extract the order parameter $M^\alpha$, which is defined as $M^\alpha=\sqrt{S^\alpha\left(\vect{Q}\right)}$, where $\vect{Q}$ is the
ordering vector associated with a peak in $S^\alpha\left(\vect{k}\right)$. 
In the thermodynamic limit, this is the equivalent to $M_0$ from MSW theory. A comparison of both quantities can be found in Fig.~\ref{fig:M_Heis_triang}. We plot both $M^x$ and $M^z$ due to the anisotropy caused by the triplet physics mentioned at the 
beginning of this section.
Discontinuous jumps in the ED magnetizations are due to the change of the spin sector hosting the ground state, 
going from the singlet sector (characterized by $M^x = M^z$) to the triplet sector (characterized by $M^x \neq M^z$).
We observe very severe deviations between the ED data on the one side and the predictions from LSW and MSW theory on the other side: 
in particular, apart from the deviations
between $M^x$ and $M^z$, the ED data appears to be almost constant over a large $\alpha$ interval. The strong
difference between ED results on the one hand and MSW/LSW predictions on the other can also 
be attributed to very significant finite-size corrections to the ED data -- finite-size effects are particularly 
pronounced here, due to the open boundary conditions of ED clusters. 
Nonetheless, for $\alpha = 1$ the magnetization of the 30-site cluster gives $M^x  = M^z \approx 0.13$, 
lying close to recent Monte Carlo estimates \cite{Heidarian2009}. 

From the location of the peak of the structure factor one can extract the vector of predominant ordering, $\vect Q$, the $x$-component of which is plotted in Fig.~\ref{fig:Q_Heis_triang}. 
Remarkably, for the 30-site cluster the $\vect Q$ corresponding to $M^x$ (labeled as $\vect Q^x$ in the figure) indicates a transition from spiral to 
 N\'eel order at around $\alpha\approx 1.4$, which lies well below the classical threshold $\alpha =2$. On the contrary, the $\vect Q$ corresponding to $M^z$ (labeled as $\vect Q^z$) increases smoothly up to $\alpha\approx 2$, where it undergoes a discontinuous transition to the square-lattice N\'eel value as well. 
However, increasing the system size from 24 to 30 spins shifts significantly the curves of $\vect Q^x$  and $\vect Q^z$  to the left, suggesting that 
for even larger sizes both curves might exhibit a discontinuous transition to the N\'eel ordering vector for a value of $\alpha$ close to 
the transition indicated by MSW, $\alpha \approx 1.3$.
Finally, we notice that at $\alpha=1$ the ED results deviate from the isotropic value $Q_x=120^\circ$ because the required threefold symmetry is broken by the shape of the simulation cluster, Fig.~\ref{fig:systemsexactdiag}.

The nearest-neighbor spin--spin correlations $T_{ij}$, Eq.~\eqref{totalspinTij},\footnote{For ED we report the values of $T_{ij}$ averaged over the central spins, where boundary effects are minimal.} are in qualitative agreement with the MSW results as well (Fig.~\ref{fig:SS_Heis_triang}). In particular, they show 1D-like behavior at small $\alpha$, a spiral phase in an intermediate parameter range around the isotropic limit $\alpha=1$, and a 2D-N\'{e}el structure at large $\alpha$. 

 Finally, we focus on the chirality correlations. Comparing such correlations for the 14, 24, and 30 spin clusters shows that they are strongly suppressed for $\alpha\lesssim 0.5$ and for $\alpha \gtrsim 1.4$ when going to larger lattice sites. This indicates that a non-spiral phase appears in this region in the thermodynamic limit, in agreement with our MSW calculations.  The persistance of significant correlations in the region $0.5\lesssim\alpha\lesssim 1.4$ indicates that spiral order in the ground state might persist in a portion of this parameter range. 
 
 In summary, despite the significant deviations in the magnitude of the order parameter, both ED and MSW theory give a coherent picture,
 both qualitatively and quantitatively,
 of the evolution of the nature of spin-spin correlations upon increasing the $\alpha$ parameter, going from quasi-1D to spiral to N\'eel.

\subsection{Discussion}
Despite its limitations, the MSW approach with ordering vector optimization reproduces faithfully the main characteristics of the phase diagram as sketched in Fig.~\ref{fig:phasediagtriang}~(b), and thus remarkably improves on the results that were previously obtained for this model with conventional spin-wave theories.
A breakdown of magnetic order -- along with a variety of observables like the ordering vector or nearest-neighbor spin--spin correlations -- indicates that
a 1D-like spin liquid might be attained below $\alpha\approx 0.65$. Due to the partial account of quantum fluctuations provided by MSW theory, we can safely
take this as a lower bound for a spin liquid in the true ground state.
Furthermore, we find a relatively small region with spiral LRO between $0.65\lesssim\alpha\lesssim 1.14$. For $\alpha\gtrsim 1.30$ the system is ordered at the 2D-N\'{e}el wave-vector. 
Between $1.14\lesssim\alpha\lesssim 1.30$ the breakdown of convergence suggests another candidate region for spin-liquid behavior.

%%%%%%%%%%%%%%%%%%%%%%%%%%%%%%%%%%%%%%%%%%%%%%%%%%%%%%%

\section{\label{cha:J1J2J3}MSW theory on the $J_1J_2J_3$ model}

\begin{figure}
        \center
%        \hspace{1.5cm}
        \includegraphics[width=0.3\textwidth]{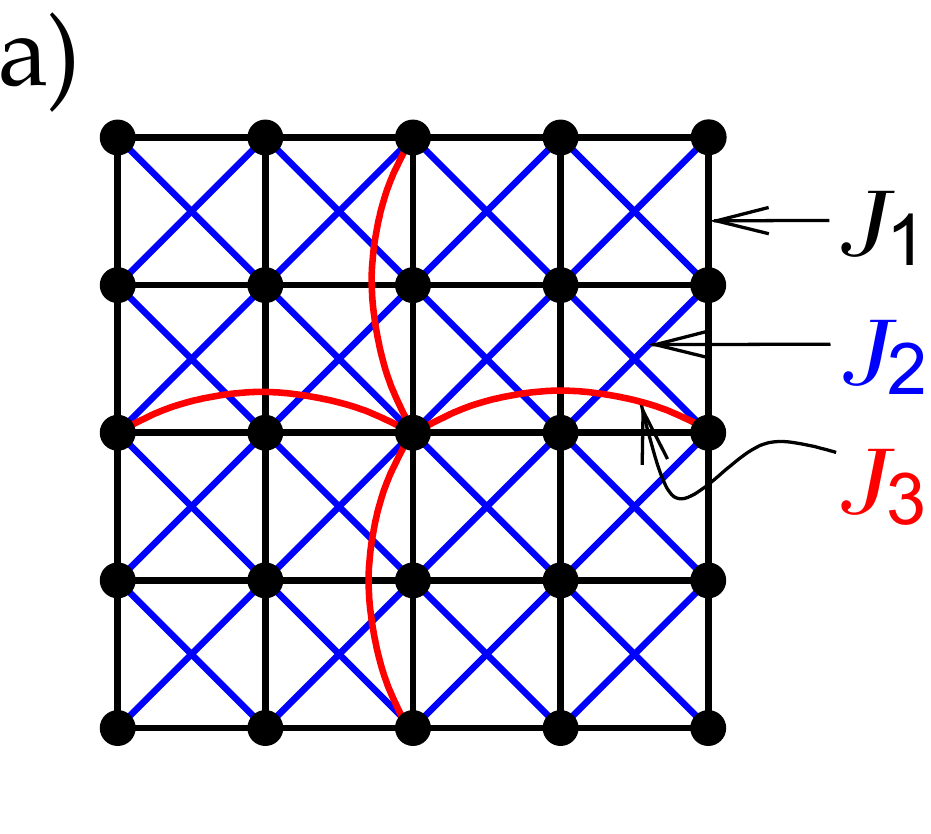}\quad\includegraphics[width=0.313\textwidth]{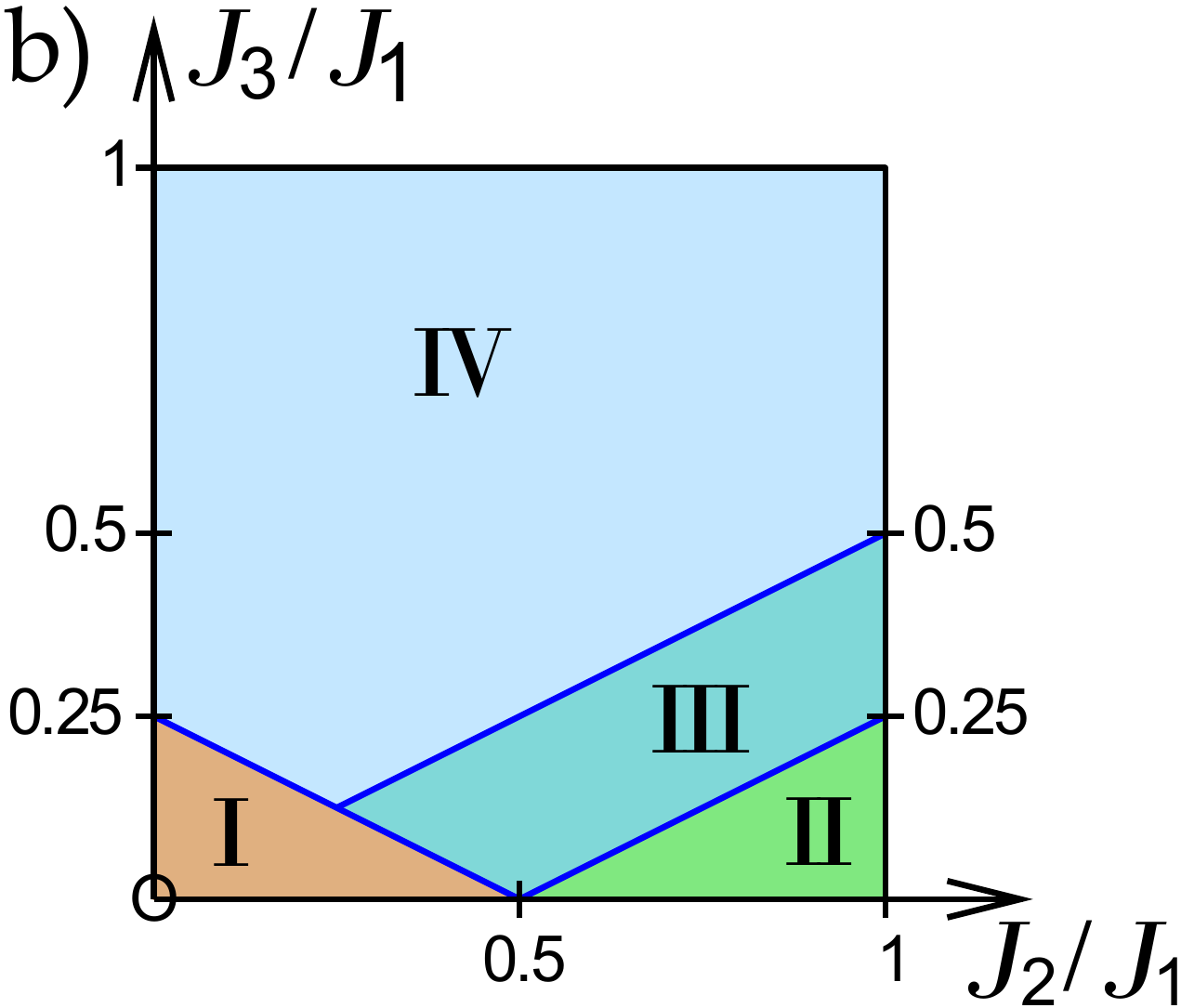}
        \caption[Geometry of the lattice in the $J_1J_2J_3$ model]{(a) A detail of the geometry of the $J_1J_2J_3$ model on a square lattice. Nearest neighbors are coupled with bonds of strength $J_1$ (black), next-nearest neighbors (along the diagonals) with $J_2$ (blue) and next-next-nearest neighbors with $J_3$ (red). 
(b) The classical phase diagram of the $J_1J_2J_3$ model shows four ordered phases: Phase I is characterized by N\'{e}el order on the square lattice. In phase II the system decouples into two independently N\'{e}el ordered sublattices with a doubled unit cell each. Phases III and IV are spirally ordered with $\vect{Q}=\left(q,\pi\right)$ and $\vect{Q}=\left(q,q\right)$, respectively.
}
        \label{fig:J1J2J3geom}
    \end{figure}

In this section, we investigate another paradigmatic frustrated spin model, the $J_1J_2J_3$ model on the square lattice. It involves couplings between nearest-neighbors (NN), $J_1$, next-nearest-neighbors (NNN), $J_2$, and next-next-nearest-neighbors (NNNN), $J_3$. 
A sketch of the geometry of the system may be found in Fig.~\ref{fig:J1J2J3geom}~(a). This model allows to continuously tune the Hamiltonian from an unfrustrated antiferromagnetic square lattice to a highly frustrated magnet.

\subsection{Classical and quantum mechanical phase diagram of the $J_1J_2J_3$ model at $T=0$}

The classical phase diagram of the $J_1 J_2 J_3$ model \cite{Gelfand1989,Moreo1990,Chubukov1991,Ferrer1993} is sketched in Fig.~\ref{fig:J1J2J3geom}~(b).
One identifies: 
\begin{itemize}
	\item[I)] A 2D-N\'{e}el phase with $\vect{Q}=\left(\pi,\pi\right)$ just as in the unfrustrated square lattice. 
	It is delimited by the classical critical line $\left(J_2+2 J_3\right)/J_1=1/2$;
	\item[II)] A phase where the system decouples into two independent $J_2-$sublattices with a doubled unit cell. Both are N\'{e}el ordered individually. This phase is infinitely degenerate because the two sublattices can be rotated one with respect to the other without affecting the energy;
	\item[III)] A spiral phase with ordering vector $\vect{Q}=\left(q,\pi\right)$, where $q$ varies continuously over the phase diagram;
	\item[IV)] A second spiral phase, this time with ordering vector $\vect{Q}=\left(q,q\right)$; $q\to\pi/2$ for $J_3\to\infty$, attaining the limit
	of two decoupled and N\'eel-ordered $J_3-$sublattices. 
\end{itemize}

This phase diagram is believed to change considerably in the quantum limit \cite{Figueirido1989,Read1991,Ferrer1993,Mambrini2006}:
In phase II quantum fluctuations select the columnar ordered states with $\vect{Q}=\left(\pi,0\right)$ or $\vect{Q}=\left(0,\pi\right)$ from all the possible classical states. Furthermore, the N\'{e}el phase I increases in size considerably and N\'{e}el order persists up to the vicinity of the line $\left(J_2+J_3\right)/J_1=1/2$. In the vicinity of this line, the classical order is believed to be destabilized and to be replaced by a non-magnetic state. The controversy about the exact nature of the ground state in this highly frustrated region, however, is still not settled. In particular, it has been suggested that it could have the nature of a \emph{columnar valence bond crystal} \cite{Leung1996} with both translational and rotational broken symmetries, of a \emph{plaquette state} with no broken rotational symmetry \cite{Mambrini2006}, or of a \emph{spin liquid} with all symmetries restored \cite{Chandra1988,Locher1990,Zhong1993,Capriotti2004a,Capriotti2004b}.

In the following, we investigate the quantum model using the modified spin-wave (MSW) formalism, and compare it to recent results from projected entangled-pairs states (PEPS) calculations. The MSW lattice size is again $N=32\times 32$.
In most of parameter space, a lattice of $N=32\times 32$ spins is essentially already converged to the infinite lattice, except close to a quantum critical point. 

In Ref.~\cite{Murg2009}, some of us reported numerical calculations of the $J_1 J_2 J_3$ model based on the projected entangled-pair state (PEPS) 
variational \emph{Ansatz} 
for varying lattice sizes. In the following, we will focus on the extrapolations to the thermodynamic limit, except if stated otherwise.

We first discuss in more detail the special cases of the $J_1J_2$ model (\emph{i.e.}, $J_3=0$) and the $J_1J_3$ model (\emph{i.e.}, $J_2=0$). Both models have been studied before within the MSW formalism \cite{Barabanov1990,Xu1990,Xu1991,Ivanov1992,Gochev1994,Dotsenko1994}. On the one hand, we confirm existing results on the $J_1 J_2$ case,
for which the optimization of the ordering wave-vector returns only two possible values (corresponding to N\'eel order [$\vect{Q}=\left(\pi,\pi\right)$] or columnar  order [$\vect{Q}=\left(\pi,0\right)$ or $\vect{Q}=\left(0,\pi\right)$]), and we give further insight into the 
spin stiffness and the dimer--dimer correlation functions. 
On the other hand, we analyze the $J_1 J_3$ model with optimization of the ordering wavevector, which proves crucial to correctly capture the quantum effects on the classical spiraling phases appearing in this case \cite{Xu1991}.  
Finally, we give an overview of the entire quantum ground state phase diagram of the $J_1J_2J_3$ model.

\subsection{Ground state properties of the $J_1J_2$ model\label{cha:J1J2}}

    \begin{figure}
        \center
        \includegraphics[width=0.75\textwidth]{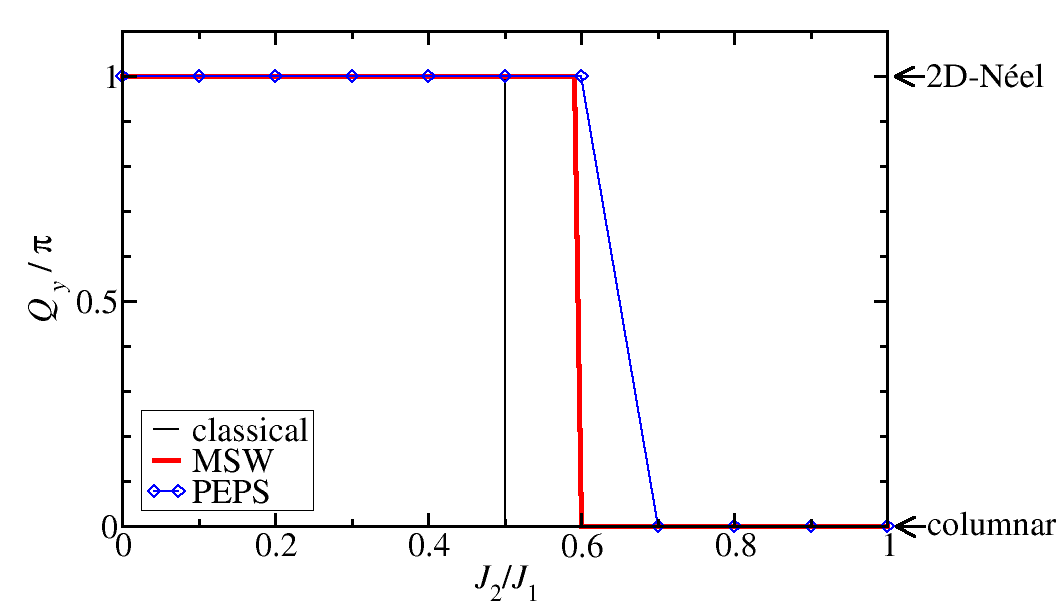}
        \caption[$y$-component of the ordering vector]{For the $J_1J_2$ model the $y$-component of the ordering vector shows a considerable shift in the quantum model with respect to the classical value.}
        \label{fig:QyJ1J2}
    \end{figure}

\begin{figure*}
				\centering
        \includegraphics[width=0.48\textwidth]{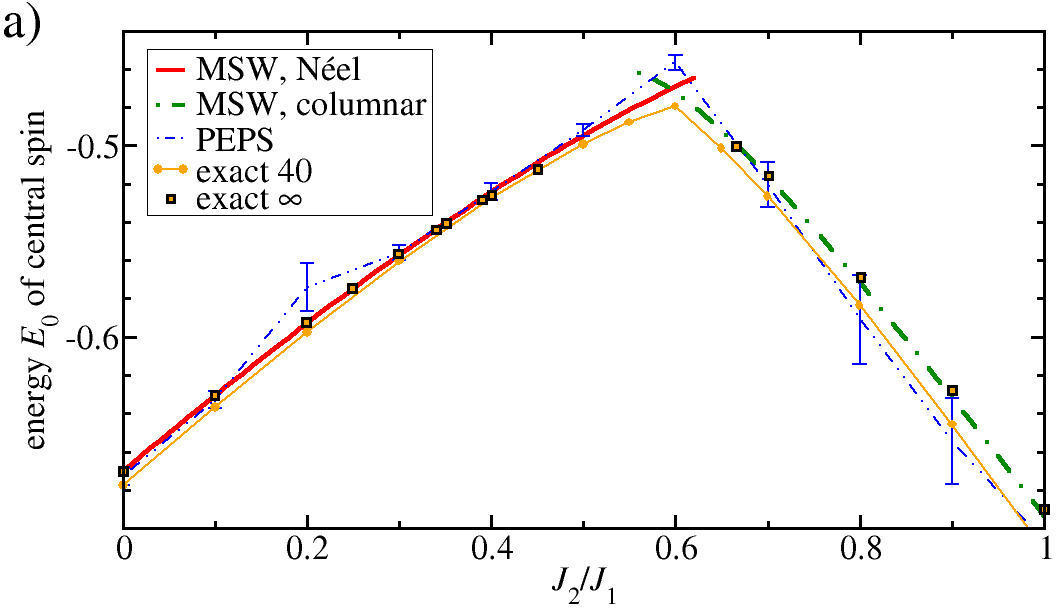}\quad\includegraphics[width=0.48\textwidth]{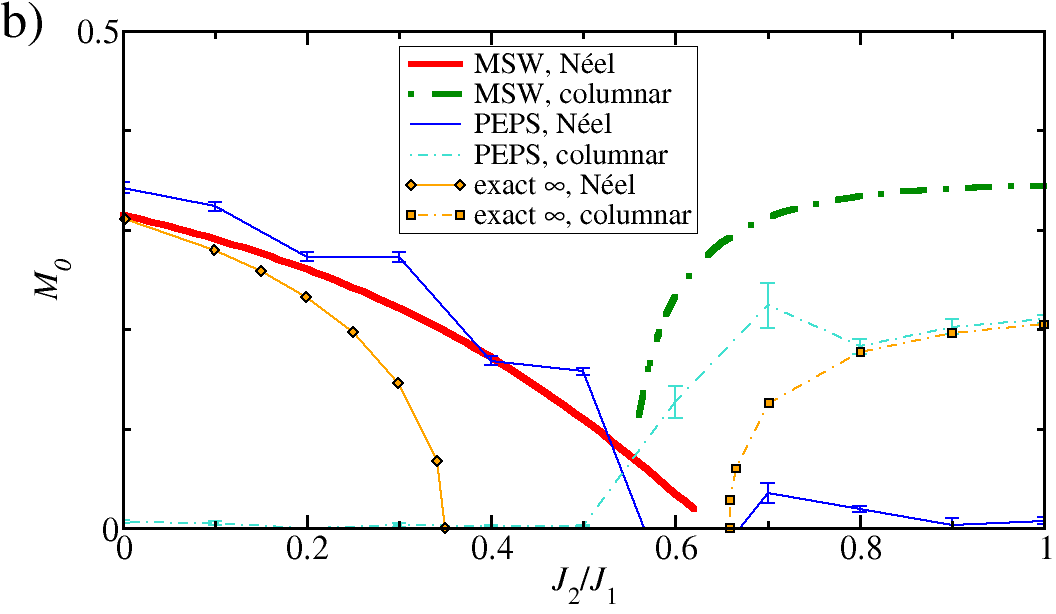}
        
				\includegraphics[width=0.48\textwidth]{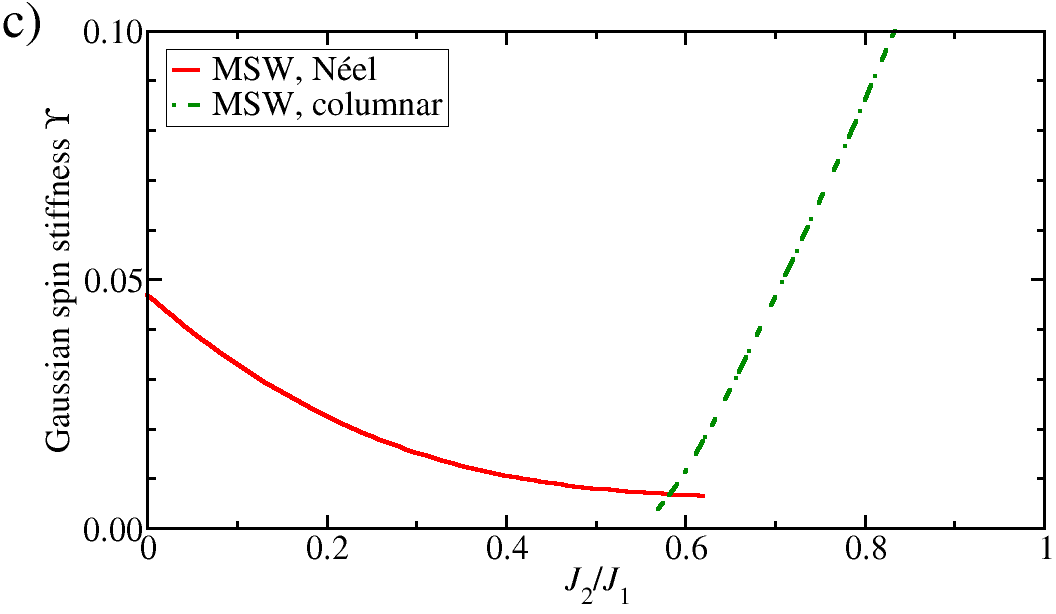}\quad\includegraphics[width=0.48\textwidth]{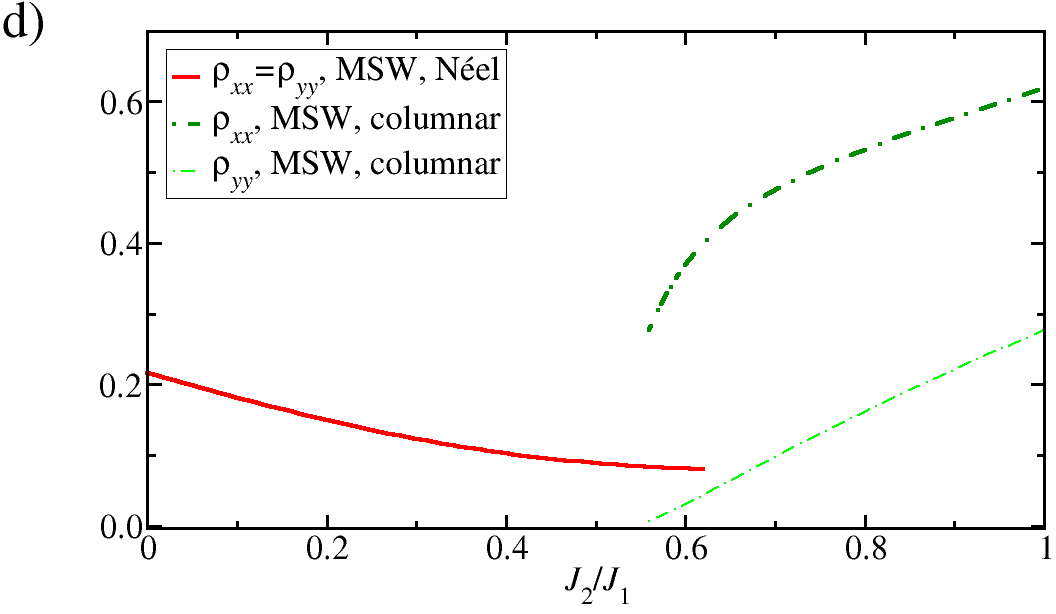}
        \caption{\label{fig:allJ1J2}
        Comparison, for the $J_1J_2$ model, of MSW data to PEPS results extrapolated to the thermodynamic limit and ED from Ref.~\cite{Richter2010} [for the 40-spin cluster (labeled `exact 40') and extrapolated to the thermodynamic limit (`exact $\infty$')]. For the MSW method, the curves obtained when starting the self-consistent iteration from a N\'{e}el state (thick red line) and from a columnar ordered state (thick dot-dashed green line) are both included.         
        The figures show
        (a) the ground state energy of the central spin;
        (b) the MSW order parameter $M_0$ compared to $M\left(\pi,\pi\right)$ (N\'{e}el) and $M\left(\pi,0\right)$ (columnar) derived from PEPS calculations;
        (c) the Gaussian spin stiffness, and (d) components of the spin stiffness tensor. In the N\'{e}el phase $\rho_{xx}=\rho_{yy}$ by symmetry. The partial spin stiffnesses $\rho_{\alpha \beta}^{\mathrm{partial}}$ are found to equal the total ones, $\rho_{\alpha \beta}$.        
        }
\end{figure*}
Figures~\ref{fig:QyJ1J2}~and~\ref{fig:allJ1J2} report the results for the $J_1J_2$ model from the MSW method as well as from PEPS calculations. For comparison, we also plot the values for the energy and magnetization that where obtained in Ref.~\cite{Richter2010} from diagonalization of small clusters.
In agreement with other methods, \emph{e.g.}, exact diagonalization (ED) \cite{Einarsson1995,Schulz1996,Richter2010} or Schwinger bosons \cite{Trumper1997}, MSW theory finds N\'{e}el order with $\vect{Q}=\left(\pi,\pi\right)$ at small $J_2/J_1$ and columnar order with $\vect{Q}=\left(\pi,0\right)$ or $\vect{Q}=\left(0,\pi\right)$ at large $J_2/J_1$ (see Fig.~\ref{fig:QyJ1J2}).
As it is well known from previous studies, there is a region between $0.56\lesssim J_2/J_1\lesssim 0.62$ where the 2D-N\'{e}el ordered and the columnar state are both stable solutions within MSW theory. The starting point of the self-consistent calculations determines which type of order is returned as the solution. However, the solutions differ in energy and therefore one of them is only a local free energy minimum of the self-consistent equations. 
The transition from 2D-N\'{e}el order to columnar order takes place at $J_2/J_1\simeq 0.6$. 
For the PEPS results, we extract the wave vector of dominant spin correlations $\vect{Q}^{\mathrm{PEPS}}$ from the location of the peak of the static structure factor,  
\begin{equation}
\label{MPEPS}
M\left(\vect{k}\right)=\sqrt{\frac 1 {N^2} \sum_{ij} \braket{\vect{S}_i\cdot\vect{S}_{j} \ue^{-i\vect{k}\cdot\vect{r}_{ij}} }}\,.
\end{equation}
In agreement with the MSW prediction, $\vect{Q}^{\mathrm{PEPS}}$ is located at the N\'{e}el value $\left(\pi,\pi \right)$ up to $J_2/J_1=0.6$, while above this it lies at the value of columnar order $\left(\pi,0 \right)$.

We find a remarkable correspondence of the ground state energy per spin between the MSW prediction and ED results extrapolated to the infinite lattice from Ref.~\cite{Richter2010} [Fig.~\ref{fig:allJ1J2}~(a)]. Moreover, the noticeable kink associated with the N\'eel-to-columnar transition of MSW theory at $J_2/J_1=0.6$ is exhibited as well by the 40-sites system from Ref.~\cite{Richter2010}. 
Therefore, ED confirms that $J_2/J_1 = 0.6$ marks a transition point, although
in the true ground state such a transition might connect the columnar state to a quantum-disordered state.
A similarly good agreement is found with the PEPS results extrapolated to the infinite size limit. 

As shown in Fig.~\ref{fig:allJ1J2}~(b), at small $J_2/J_1$, \emph{i.e.}, deep in the N\'{e}el phase, the finite size extrapolation of the ED staggered magnetization from Ref.~\cite{Richter2010} lies very close to the MSW results. 
As it is well known \cite{Takahashi1989}, in the unfrustrated square lattice limit ($J_2=0$) the MSW value $M_0=0.303$ is only slightly smaller than  $M=0.311$ from ED.
For the PEPS calculations an analogous quantity can -- similar to section~\ref{cha:ED} -- be derived from the peak height of the static structure factor, Eq.~\eqref{MPEPS}. We show its finite size extrapolation in Fig.~\ref{fig:allJ1J2}~(b). In the N\'eel phase PEPS agrees very well with MSW theory, considerably better than ED, which decreases faster towards the strongly frustrated region.
In the entire columnar phase, however, PEPS and ED data lie closer together, while MSW overestimates the order parameter. 
Around the transition, however, agreement between PEPS and MSW theory is very good. The PEPS data suggest that the magnetically disordered region, predicted by ED to occur in the range $0.35\lesssim J_2/J_1\lesssim0.66$, is either much smaller or does not occur at all. 

The MSW spin stiffness $\rho_{\|}\equiv\left(\rho_{xx}+\rho_{yy}\right)/2$, however, while being finite for any considered value of the ratio $J_2/J_1$, is strongly suppressed in the region $0.3 \lesssim J_2/J_1 \lesssim 0.6$ [Figs.~\ref{fig:allJ1J2}~(c) and~(d)], suggesting as usual that accounting for quantum fluctuations beyond the MSW approximation could lead to the disappearance of magnetic order. 
A suppression of spin stiffness is also observed in previous results coming from ED of finite clusters \cite{Einarsson1995} or from the Schwinger boson approach \cite{Trumper1997,Manuel1998}. As a consequence, even though MSW admits a stable solution with magnetic order for any $J_2/J_1$ value, for $J_2/J_1=0.6$ it exhibits a clear transition from \emph{soft} N\'eel order to a \emph{stiff} columnar order, suggesting that this transition could actually separate the columnar state from a quantum disordered phase.

\subsubsection{Dimer correlations in the  $J_1 J_2$ model.}
\begin{figure}
        \centering
        \includegraphics[width=0.225\textwidth]{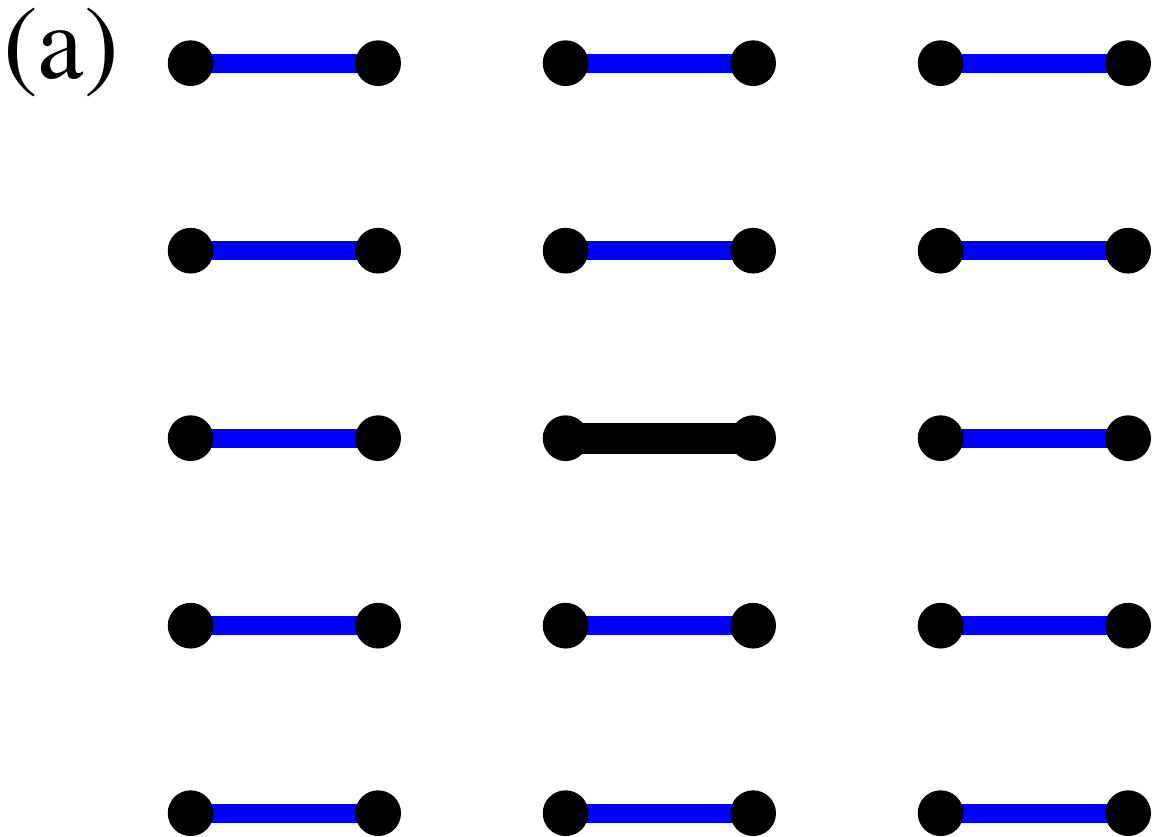}\qquad\includegraphics[width=0.225\textwidth]{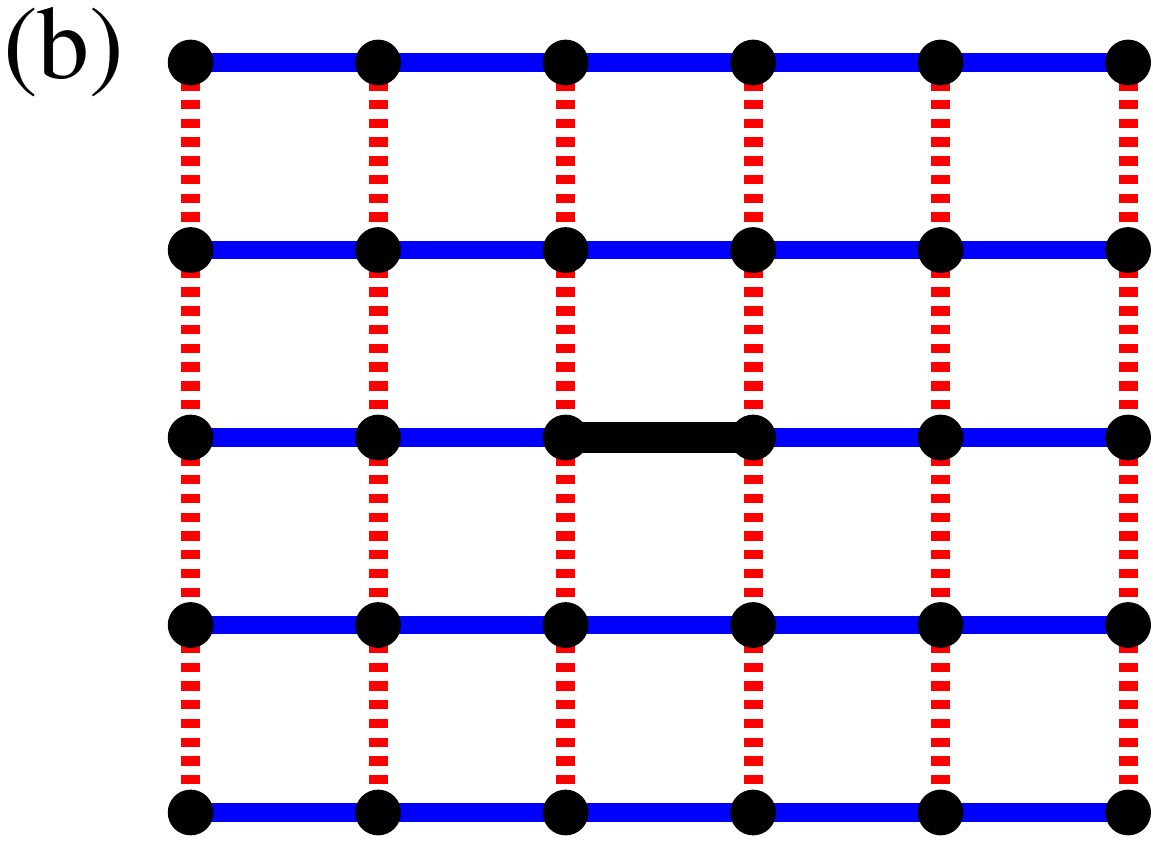}
        \caption{Sketch of dimer--dimer correlations in (a) a valence bond crystal and (b) a columnar state. Black dots are lattice sites. Blue solid (red dashed) lines are dimers correlated (anti-correlated) with the central dimer (thick black line).}
        \label{fig:sketchdimerstates}
    \end{figure}

The nature of the state in the transition region between N\'{e}el and columnar order, where magnetic order is strongly reduced, can be further investigated through the study of the dimer--dimer correlations 
\begin{equation}
    C_{ijkl}=\braket{\left(\vect{S}_i\cdot\vect{S}_j\right)\left(\vect{S}_k\cdot\vect{S}_l\right)}\,,
\end{equation}
where $k$ and $l$, and $i$ and $j$ are pairs of neighboring spins. 
Figure~\ref{fig:sketchdimerstates} sketches the expectation for the dimer--dimer correlations in (a) a columnar valence bond crystal and (b) a columnar magnetic state.

\begin{figure}
        \center
%        \hspace*{-0.01\textwidth}
        \hspace*{1cm}\includegraphics[width=0.6\textwidth]{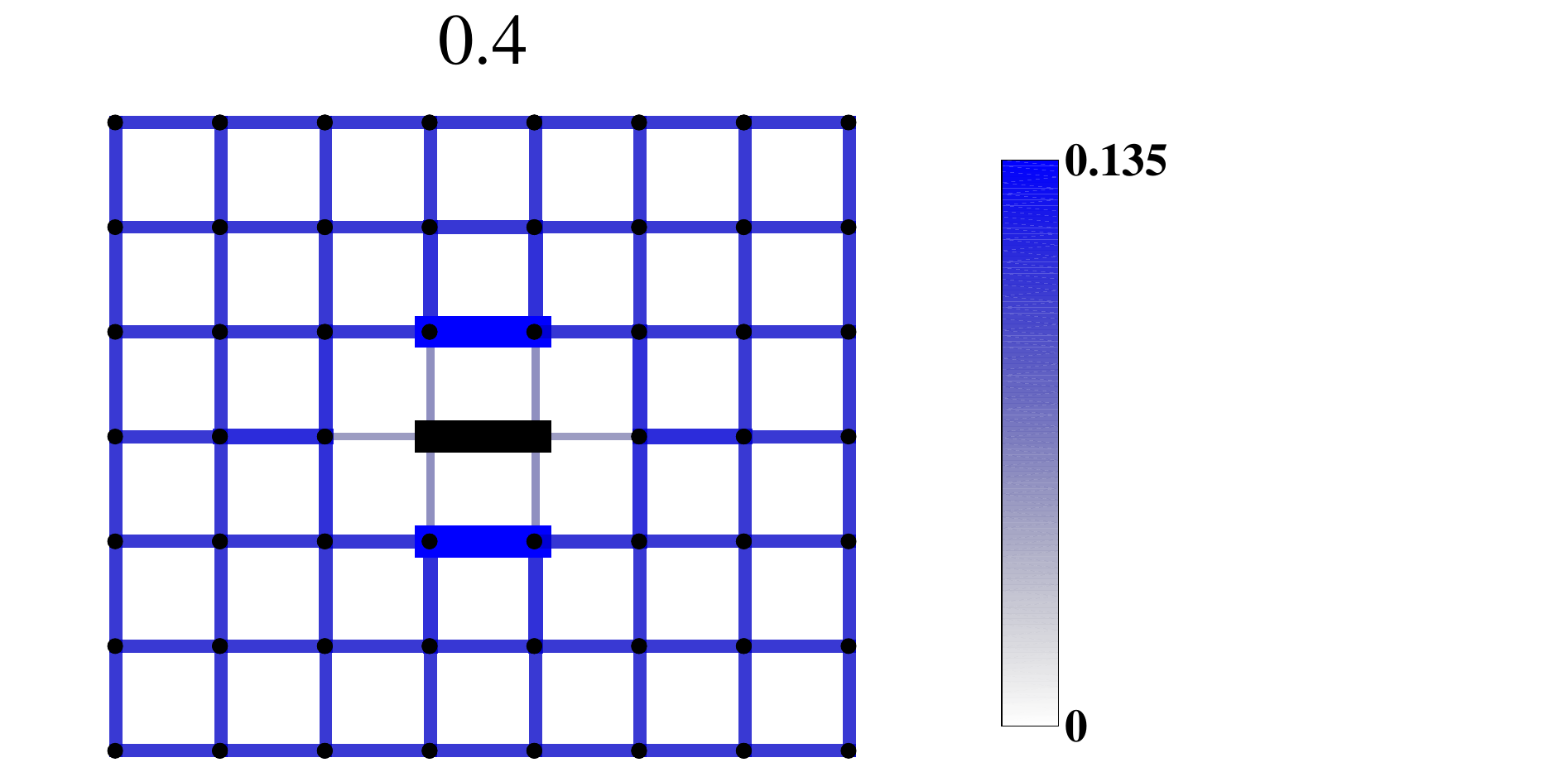}\hspace*{-1.8cm}\includegraphics[width=0.6\textwidth]{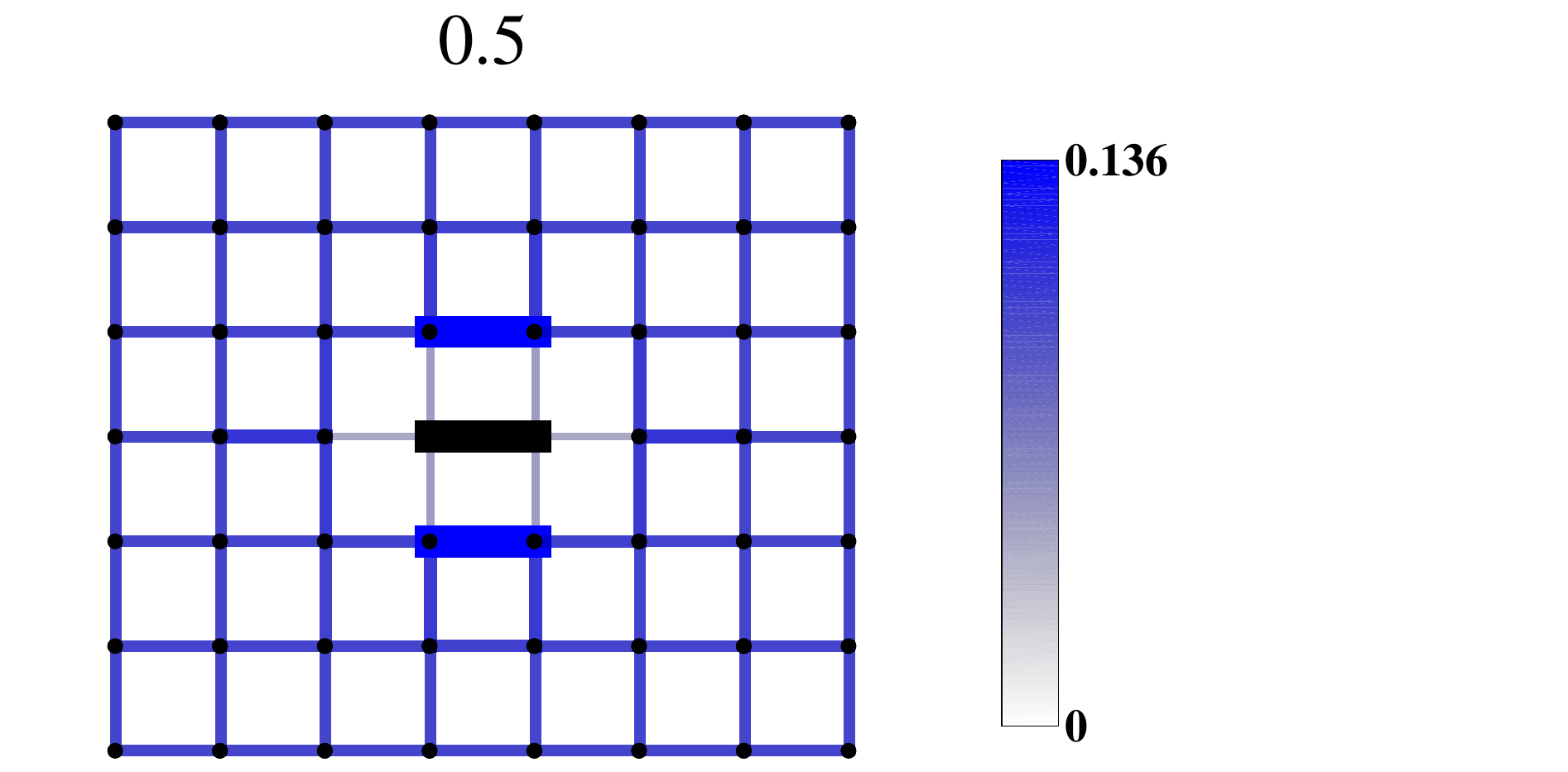}\\
        \hspace*{1cm}\includegraphics[width=0.6\textwidth]{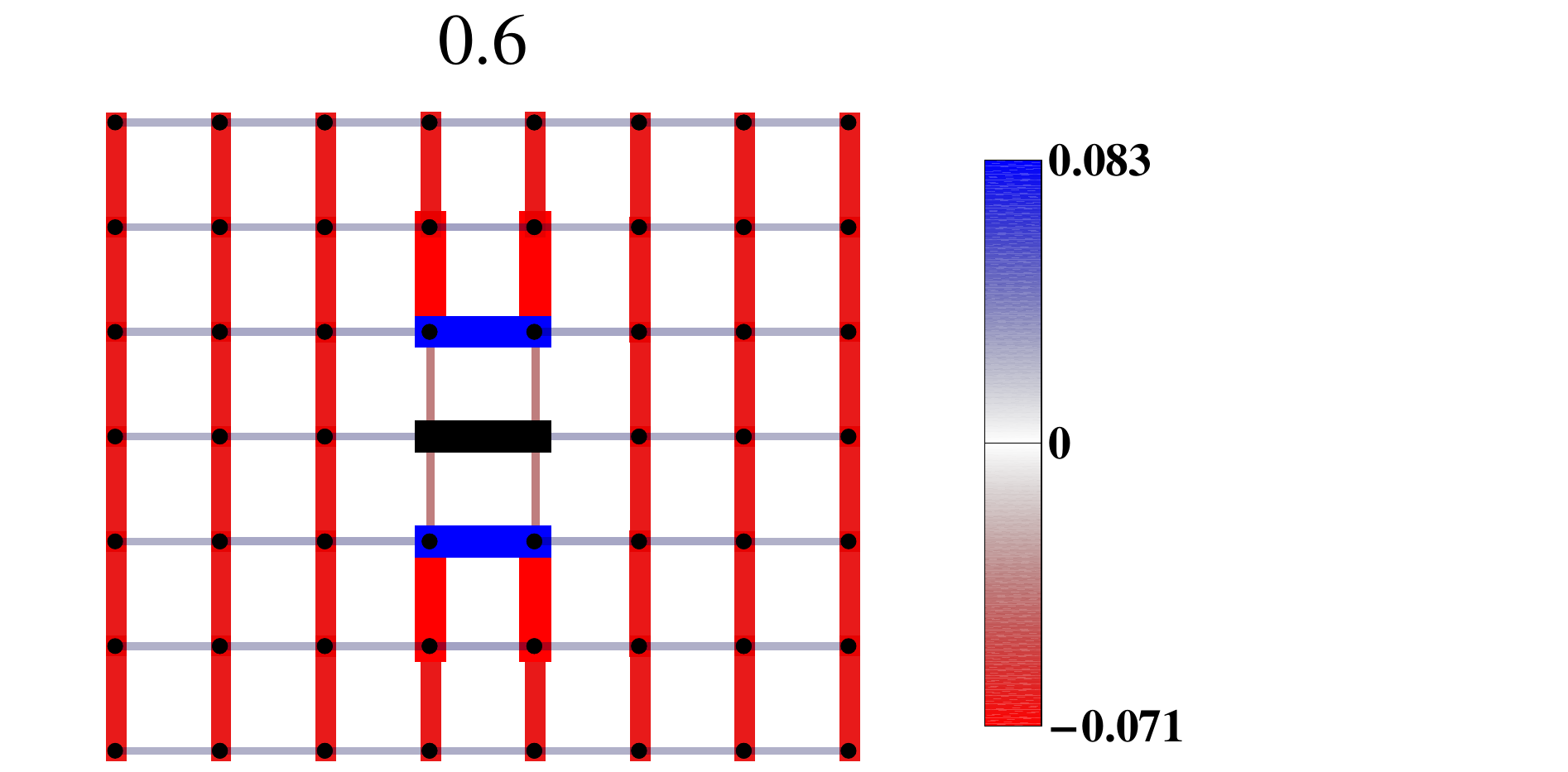}\hspace*{-1.8cm}\includegraphics[width=0.6\textwidth]{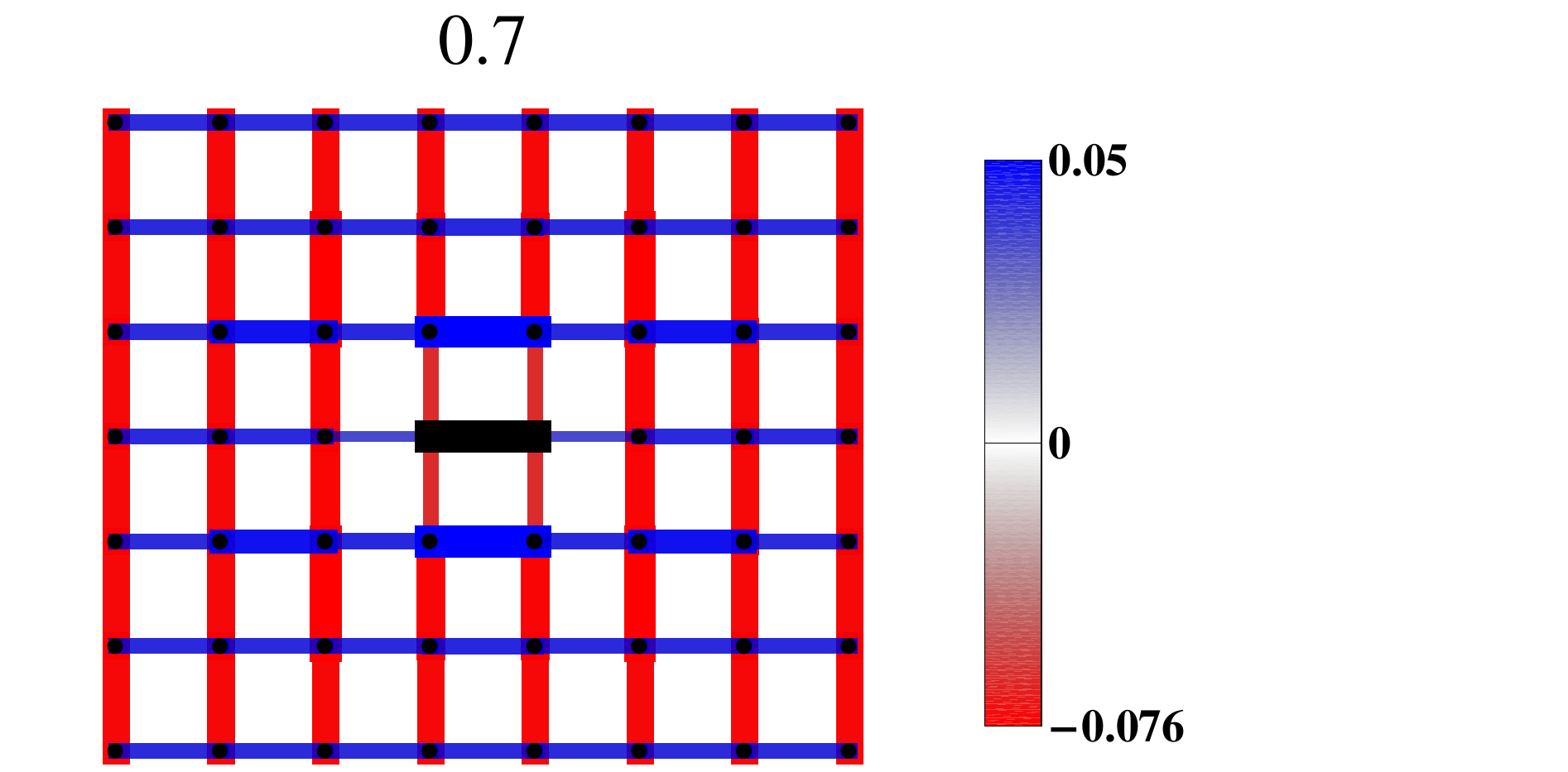}\\
        \hspace*{1cm}\includegraphics[width=0.6\textwidth]{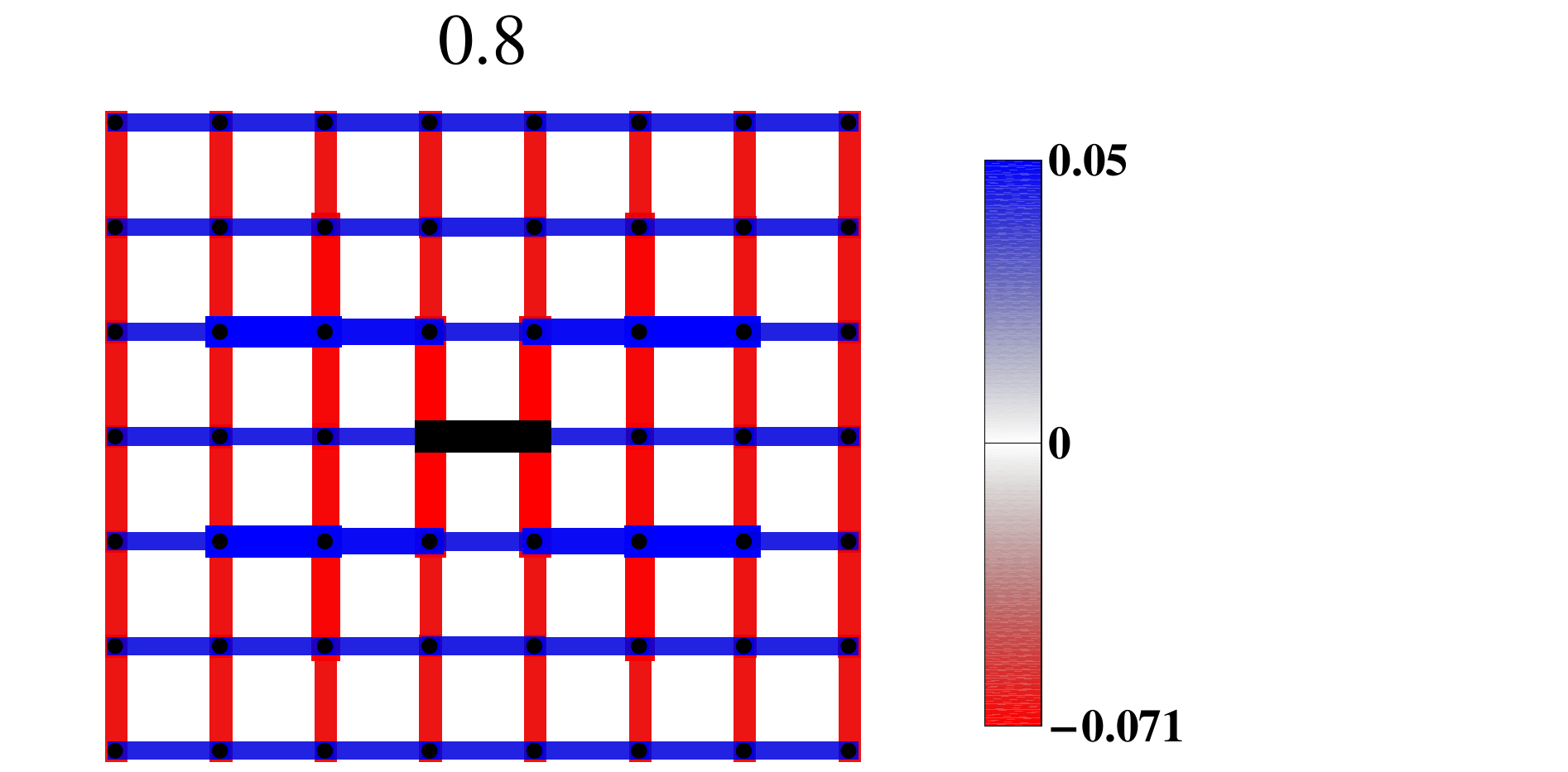}\hspace*{-1.8cm}\includegraphics[width=0.6\textwidth]{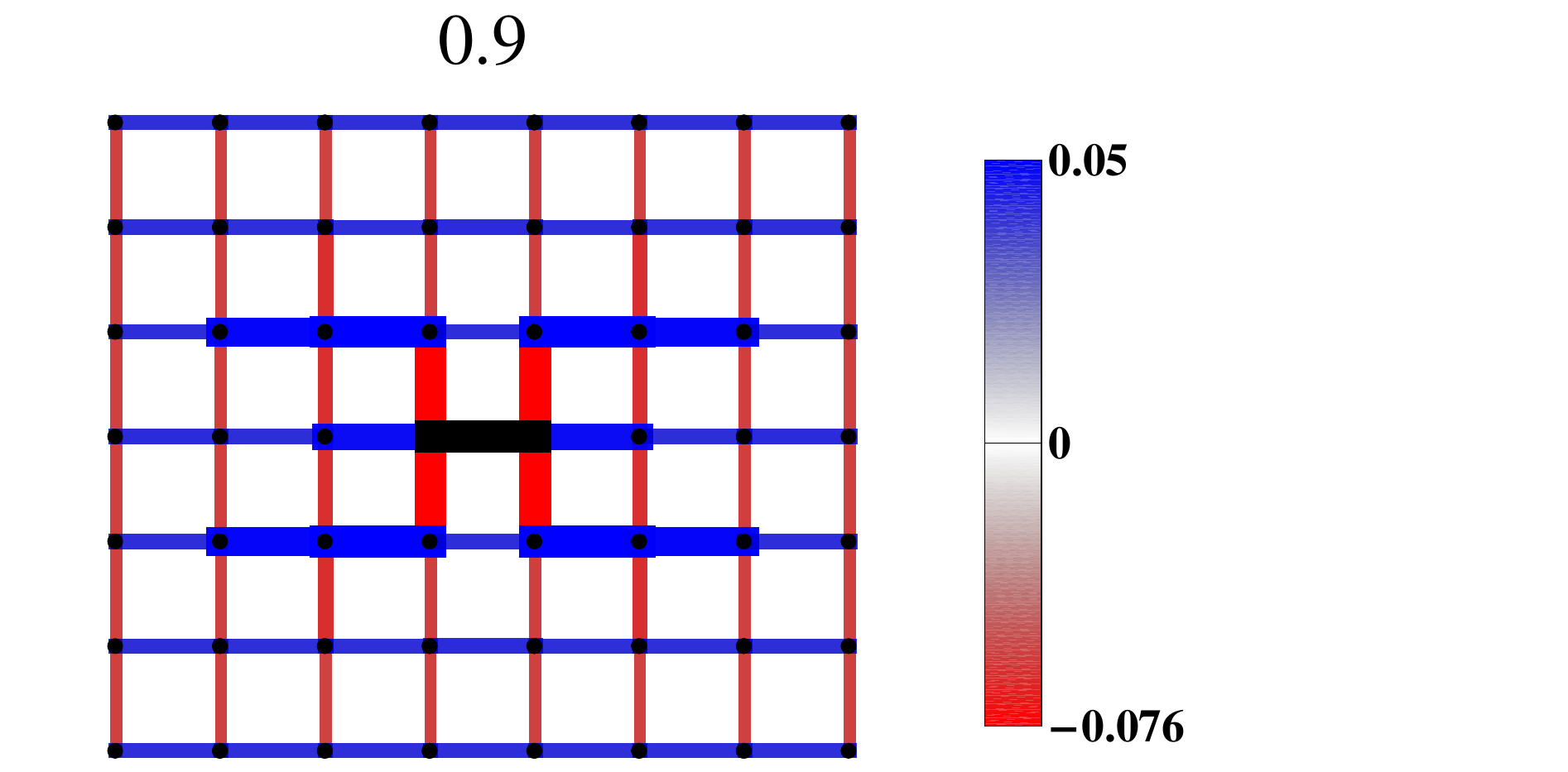}\\
        \caption{MSW correlations of the black central dimer with the other dimers of a 32x32 lattice (zoom on central region). The thickness of the lines is a non-linear function of the absolute strength of the dimer correlations. Note the change of the maximum of the linear color-scales for different values of $J_2/J_1$. Below $J_2/J_1=0.4$ and above $J_2/J_1=0.9$, the qualitative changes are minimal.
				}
        \label{fig:J1J2_dimdim}
    \end{figure}

In Fig.~\ref{fig:J1J2_dimdim}, we show the spatially resolved dimer--dimer correlations from MSW theory. Below $J_2/J_1=0.6$ the dimer-dimer correlations have a structure
compatible with a N\'eel state (namely they are positive and nearly equal for all bond pairs), while above $J_2/J_1=0.6$ the dimer--dimer correlations acquire the 
expected structure in a columnar state, with opposite signs for the correlations between dimers of the same spatial orientation (both horizontal and both vertical)
and between dimers of opposite orientations. Nonetheless, 
for $J_2/J_1\lesssim 0.7$, remarkably MSW theory shows a short-range modulation in the strength of the dimer correlations whose structure is compatible
with that of a valence bond crystal. Although MSW theory is not appropriate to characterize non-magnetic states such as a valence bond crystal, 
it is remarkable to observe that it identifies a columnar valence-bond structure as the dominant form of dimer correlations at short range;
this indication is consistent with, \emph{e.g.}, the results of PEPS \cite{Murg2008}, which also point towards columnar valence-bond
order in the non-magnetic region of the $J1J2$ model. 

\subsection{Ground state properties of the $J_1J_3$ model\label{cha:J1J3}}

\begin{figure}
        \center
%        \hspace*{-0.04\textwidth}
        \includegraphics[width=0.75\textwidth]{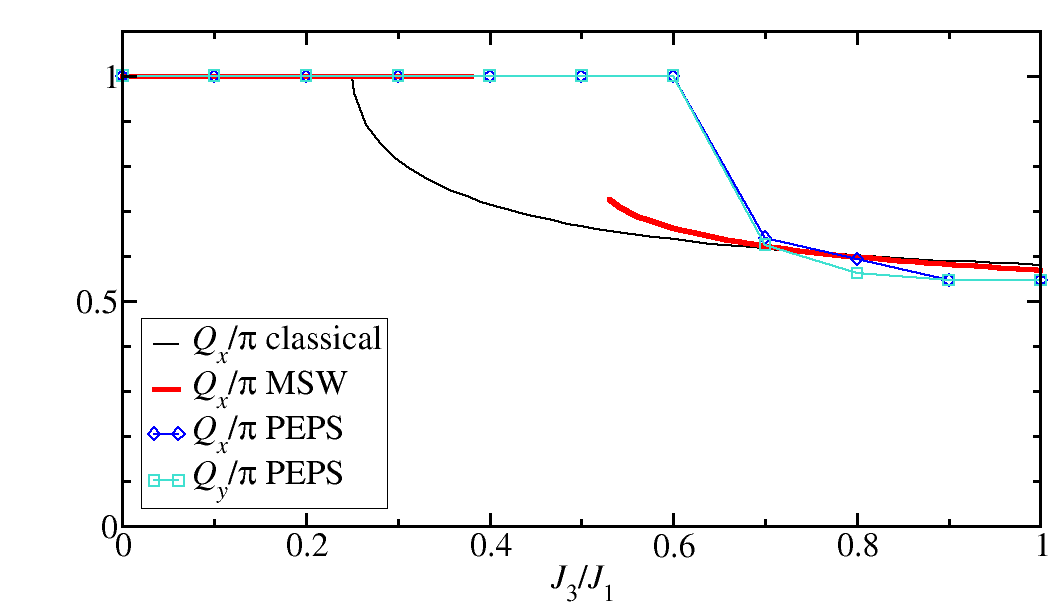}
        \caption[Both components of the ordering vector]{Position $\left(Q_x,Q_y\right)$ of the peak of the structure factor for PEPS and the ordering vector $Q_x=Q_y$ of MSW theory for the $J_1J_3$ model. A comparison to the classical ordering vector $Q_x^\mathrm{cl}=Q_y^\mathrm{cl}$ shows that quantum fluctuations stabilize N\'eel order.
        }
        \label{fig:QJ1J3}
    \end{figure}

We now turn to the $J_1J_3$ model. 
Classically, this model has a transition from N\'{e}el to spiral order at $J_3=0.25 J_1$.
Recent PEPS calculations show that for $S=1/2$ N\'{e}el order persists up to approximately $J_3/J_1=0.3$ \cite{Murg2009}. Above this point the peak of the structure factor is still at the N\'{e}el ordering vector $\left(\pi,\pi\right)$ but its height vanishes in the thermodynamic limit, which suggests a complete loss of magnetic LRO. A different type of LRO arises anew at approximately $J_3/J_1=0.6$ with an ordering vector $\vect{Q}=\left(q,q\right)$ that tends to $\left(\pi/2,\pi/2\right)$ in the limit of large $J_3$ (see Fig.~\ref{fig:QJ1J3}). For large enough $J_3$ the nature of the ordered phase becomes similar to that of the classical limit.

\begin{figure*}
        \includegraphics[width=0.48\textwidth]{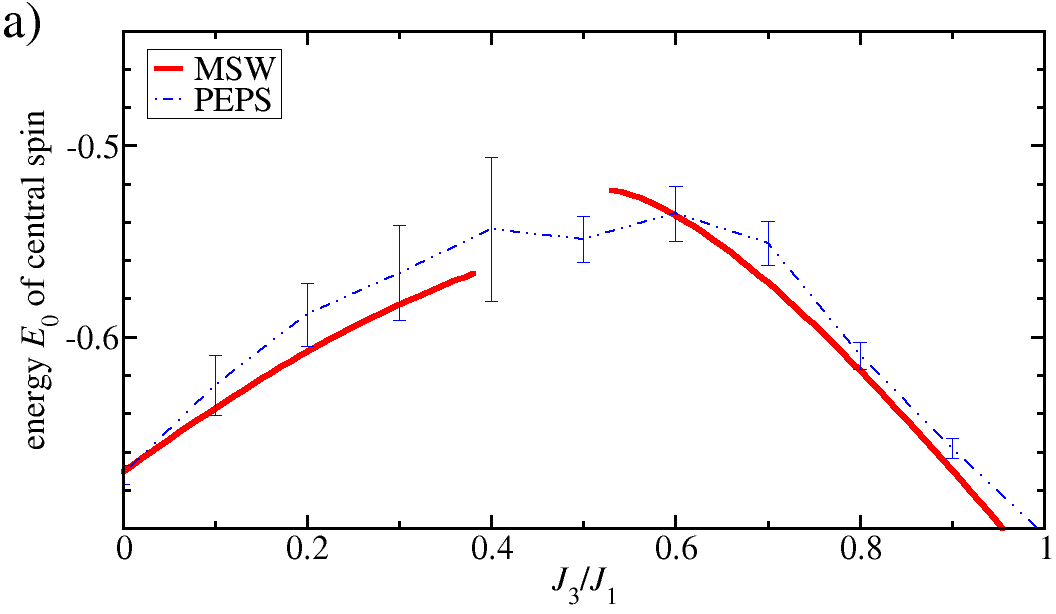}\quad\includegraphics[width=0.48\textwidth]{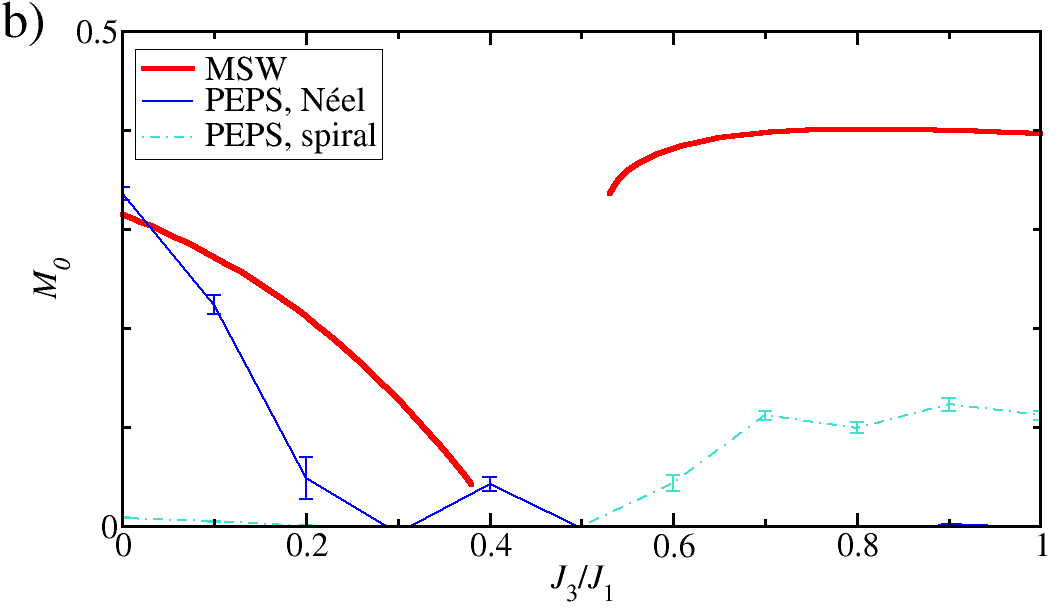}
        
				\includegraphics[width=0.48\textwidth]{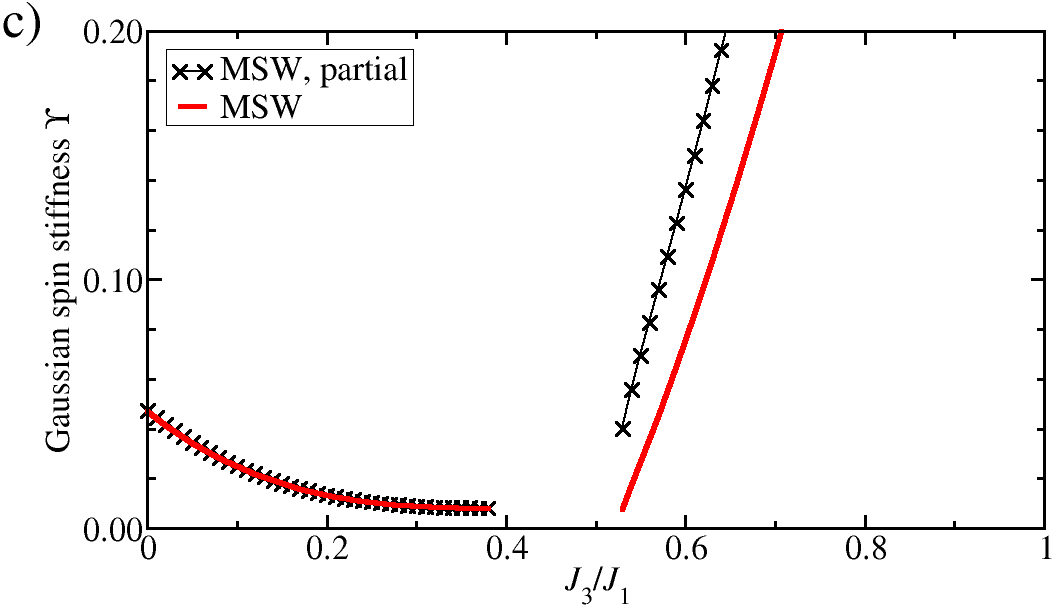}\quad\includegraphics[width=0.48\textwidth]{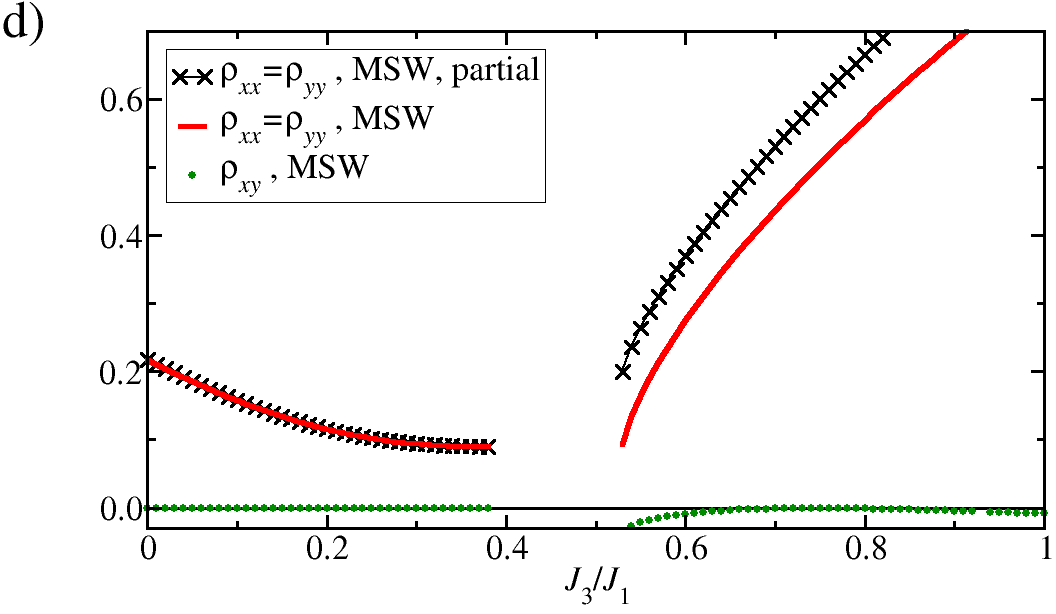}

        \caption[Comparison of MSW and PEPS results]{\label{fig:allJ1J3}
        MSW and PEPS results on the $J_1J_3$ model. Shown is (a) the energy per spin; (b) the order parameter $M_0$ for MSW theory and for PEPS the peak height of the structure factor at $\vect{Q}=\left(\pi,\pi\right)$ (N\'eel) and $\vect{Q}=\left(q,q\right)$ (spiral); (c) the Gaussian spin stiffness; (d) the components of the spin-stiffness tensor (where $\rho_{xx}=\rho_{yy}$ by symmetry, and $\rho_{xy}^{\mathrm{partial}}\equiv 0$).
        }
\end{figure*}

The optimization of the ordering wave-vector within MSW calculations shows that, for small $J_3/J_1$,  N\'eel order persists up to $J_3/J_1=0.39$
(see Fig.~\ref{fig:QJ1J3}), confirming the assumption that quantum fluctuations stabilize N\'eel order against spiral order with respect to the classical limit. 
Coming from the opposite limit of $J_3 \sim J_1$, we observe a spiral phase with continuously varying pitch vector $\vect{Q} = (q,q)$, where 
 $q$ approaches $\pi/2$ for $J_3/J_1 \to \infty$, and increases up to $q \approx 0.7 \pi$ for $J_3/J_1 \to 0.52^{+}$.
 In the region $0.39 < J_3/J_1 < 0.52$, convergence of the MSW calculations breaks down, which points at a possible spin-liquid phase, in agreement with the predictions from PEPS calculations.
 
Fig.~\ref{fig:allJ1J3}~(a) shows the PEPS energy extrapolated to the thermodynamic limit. Agreement to the MSW results is again found to be extremely good.

The indication of a disordered phase drawn from the break down of MSW theory is further corroborated by the order parameter $M_0$ [Fig.~\ref{fig:allJ1J3}~(b)], which decreases strongly for $J_3/J_1 \to 0.39^-$ and for $J_3/J_1 \to 0.52^+$, and by the spin stiffness [Fig.~\ref{fig:allJ1J3}~(c) and~(d)], which is drastically reduced when approaching the above two limits. In particular, the Gaussian spin stiffness $\Upsilon$ is already strongly reduced for $J_3/J_1\gtrsim 0.3$.
These results are consistent with the vanishing of the spin stiffness at $J_3/J_1=0.35$ that was found by ED of a system of 20 sites in Ref.~\cite{Bonca1994}.

A destabilization of magnetic order at around $J_3/J_1\gtrsim 0.3$ seems to be confirmed by the PEPS order parameter, Fig.~\ref{fig:allJ1J3}~(b), which vanishes in the range $0.3 \lesssim J_3/J_1 \lesssim 0.5$. Note that, again, we find that the PEPS order parameter deep in the N\'eel phase is similar to the MSW data, but that in the spiral phase MSW data for the order parameter lie well above the PEPS ones.

In our calculations, despite using the same equations as in Ref.\ \cite{Xu1991}, we find a considerably larger breakdown region. However, the region where our calculations do not yield a result is very stable, i.e., it does not depend much on system size nor on the exact algorithm for solving the self-consistent MSW equations. 
  
The precise nature of the state in the candidate region for quantum-disordered behavior cannot be determined reliably by the use of MSW theory. From an analysis of the dimer--dimer correlations in the convergence regions, we can find no indications of any exotic disordered quantum state; on the contrary, PEPS results indicate a plaquette state in the region of maximal frustration $J_3\approx J_1/2$ \cite{Murg2009}.

\subsection{Ground state phase diagram of the $J_1J_2J_3$ model}

\subsubsection{MSW results}

\begin{figure*}
        \centering
        \includegraphics[width=0.49\textwidth]{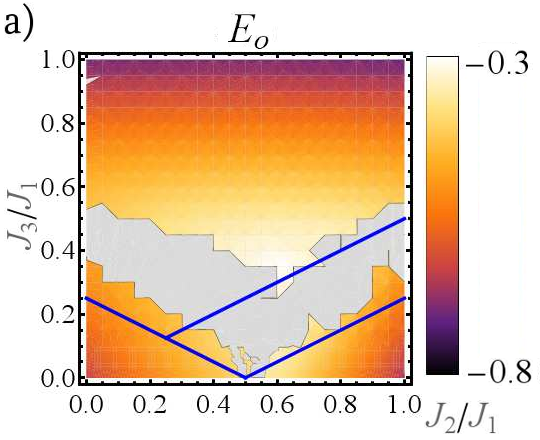}\includegraphics[width=0.49\textwidth]{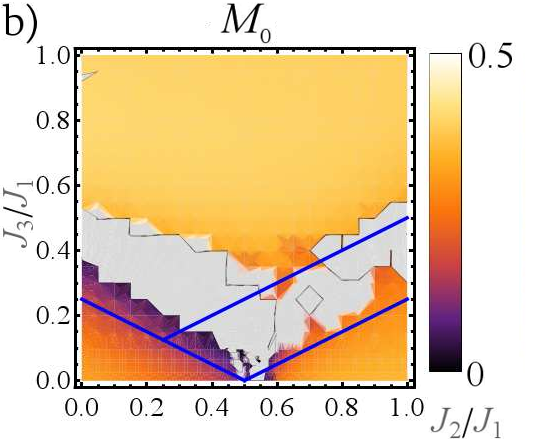}
        \vspace{0.5cm}

        \includegraphics[width=0.49\textwidth]{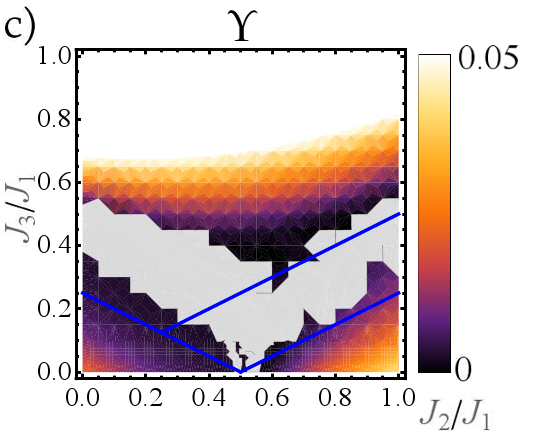}\includegraphics[width=0.49\textwidth]{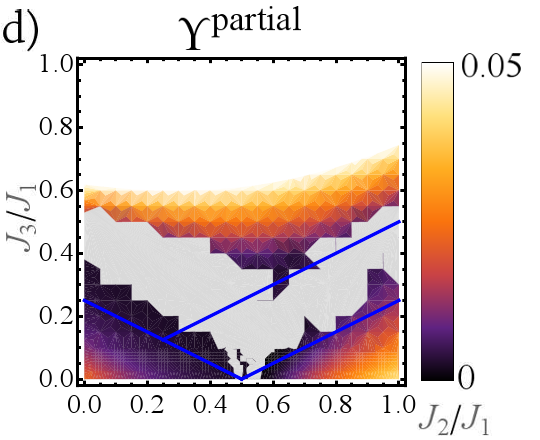}

        \caption{\label{fig:allJ1J2J3} (a) Ground state energy per spin $E_0$, (b) order parameter $M_0$, (c) Gaussian spin stiffness $\Upsilon$, and (d) Gaussian spin stiffness $\Upsilon^{\mathrm{partial}}$ calculated via Eq.~\eqref{gcpartial}. Note that $\Upsilon$ and $\Upsilon^{\mathrm{partial}}$ rise beyond the linear scale in the upper half of the plot. In the gray areas convergence of the self-consistent equations could not be reached. The blue lines are the classical phase boundaries.
        }
\end{figure*}

\begin{figure*}
        \centering
        \includegraphics[width=0.49\textwidth]{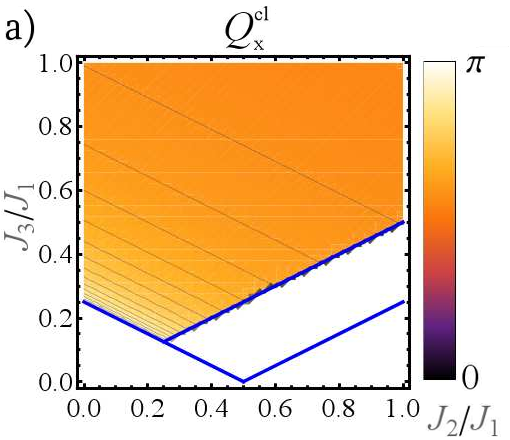}\hspace*{0cm}\includegraphics[width=0.49\textwidth]{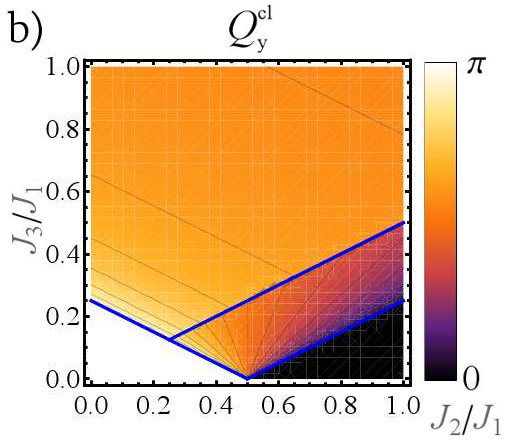}
        \vspace{0.5cm}

        \includegraphics[width=0.49\textwidth]{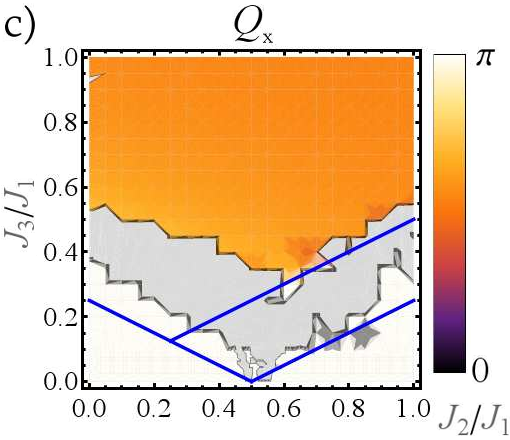}\hspace*{0cm}\includegraphics[width=0.49\textwidth]{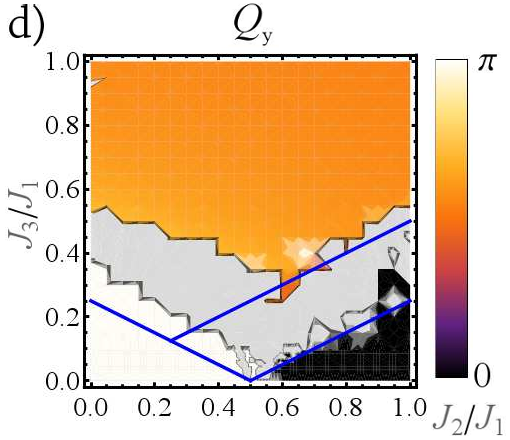}

        \caption{\label{fig:QsJ1J2J3}Ordering vector for the $J_1J_2J_3$ model in linear color scale. In the gray area convergence of the self-consistent equations could not be reached.
         (a) $x$-component, and (b) $y$-component of the classical ordering vector;
         (c) $x$-component, and (d) $y$-component of the quantum mechanical MSW ordering vector.
         The blue lines are the classical phase boundaries.
        }
\end{figure*}

After having investigated the two limiting cases of the $J_1J_2$ and the $J_1J_3$ models, we consider more generally the $J_1J_2J_3$ model over the relevant parameter range
$0 \leq J_2/J_1, J_3/J_1 \leq 1$.
As already seen in the case of the $J_1 J_3$ model, we observe a sizable parameter range over which the convergence of MSW theory breaks down, and which
is then pointed out as a candidate region for non-magnetic behavior. We notice that, while convergence is achieved for any $J_2/J_1$ ratio at $J_3 = 0$, a region
of convergence breakdown opens up by adding a small $J_3$ component around $J_2/J_1 \approx 0.5$. 
The energy per spin increases when approaching this region, showing the increased influence of frustration [Fig.~\ref{fig:allJ1J2J3}~(a)].
The indications for a quantum disordered phase in the break-down region is corroborated by the decrease of the order parameter [Fig.~\ref{fig:allJ1J2J3}~(b)] and the spin stiffness [Fig.~\ref{fig:allJ1J2J3}~(c) and~(d)] when approaching the break-down region.  
 
The nature of the phases where MSW reaches convergence can be seen in the ordering vector, which we display in Fig.~\ref{fig:QsJ1J2J3} in comparison with the classical one. We find three ordered phases: 1) For small $J_3/J_1$ and $J_2/J_1$ we find a N\'{e}el ordered phase. Its boundary is pushed upwards to higher values of $J_3/J_1$ with respect to the classical limit; 2)  a columnar phase is found at small $J_3/J_1$ but larger $J_2/J_1 \gtrsim 0.6$; 3) for large $J_3/J_1$ a spiral phase arises with an ordering vector $\vect{Q}=\left(q,q\right)$ that approaches $\vect{Q}=\left(\pi/2,\pi/2\right)$ for large $J_3/J_1$.
As a consequence, a most dramatic effect of quantum fluctuations seems to be the disappearance of phase III in the classical phase diagram, characterized
by magnetic order at a pitch vector $\vect{Q}_{cl} =\left(q,\pi\right)$ with continuously varying $q$, in favor of the columnar phase and of a potentially
quantum-disordered phase.

\subsubsection{Comparison to PEPS calculations}

\begin{figure}
        \centering
        \hspace*{-0.02\textwidth}\includegraphics[width=0.49\textwidth]{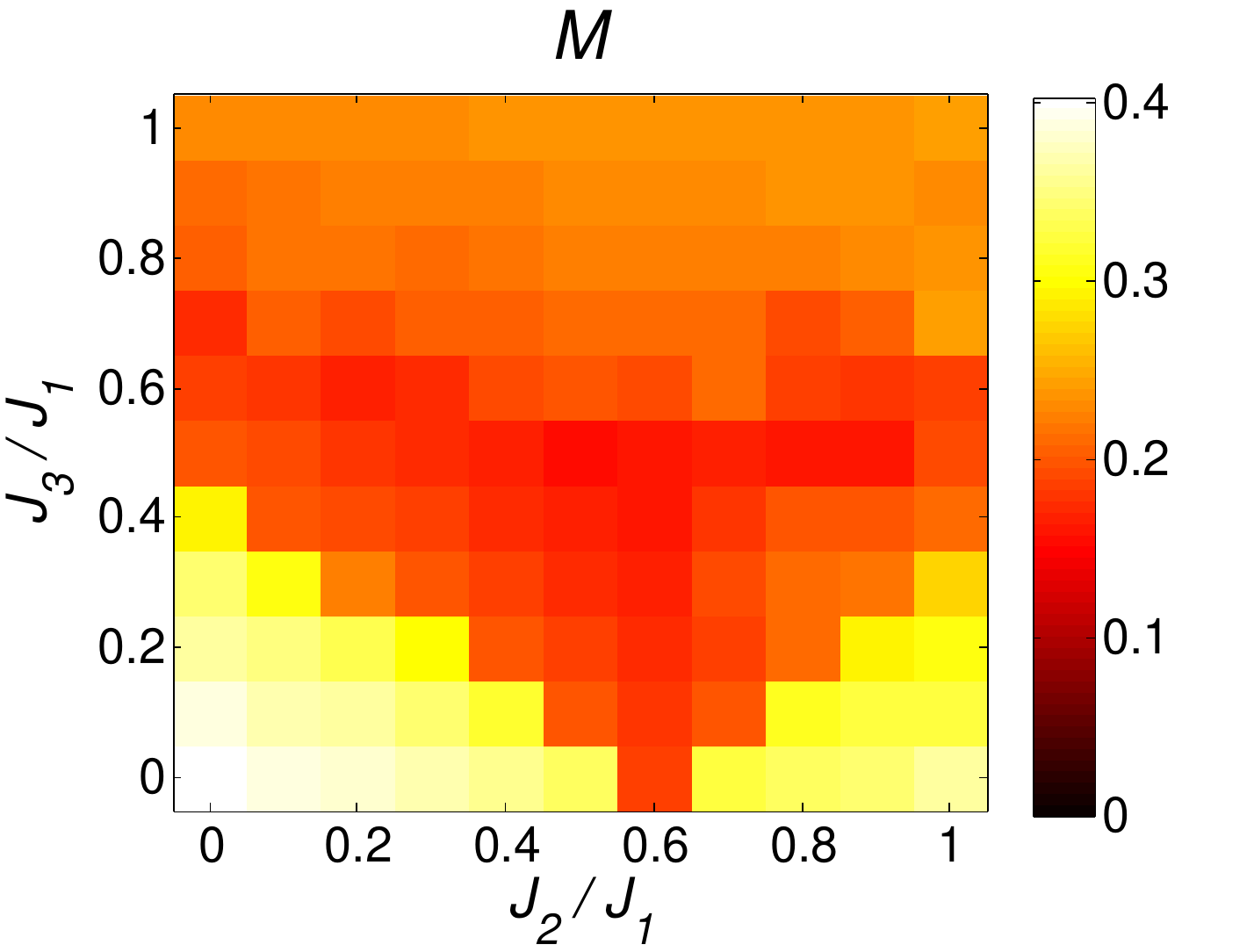}%\\
        \caption{
        $M\left(Q\right)$ [Eq.\eqref{MPEPS}] for a PEPS calculation on a $8\times 8$ lattice with auxiliary dimension $D=3$. A low value marks a destabilization of magnetic LRO.}
        \label{fig:J1J2J3PEPS_M}
    \end{figure}

In Fig.~\ref{fig:J1J2J3PEPS_M}, we display the peak height of the static structure factor, Eq.~\eqref{MPEPS}, from a PEPS calculation on a $8\times 8$ lattice with auxiliary dimension $D=3$. We observe a broad asymmetric v-shaped region in which the 
magnetic order, quantified by the height of the peak in the structure factor, is strongly suppressed. We notice that this 
region is strongly reminiscent of (albeit broader than) the breakdown region of MSW theory.
In particular, the asymmetry is due to the fact that the bottom of the ``v'' lies at $J_2/J_1 > 0.5$, a characteristic which is shared with 
the MSW phase diagram. While a thorough finite-size scaling analysis of the PEPS data would be necessary to determine
the precise boundaries of the possible magnetically disordered regions, a quantitative information can be extracted
even from the finite-size PEPS data concerning the location of the pitch vector of the dominant (long-ranged or short-ranged) 
magnetic correlations.     

Similarly to what happens in the above spin-wave calculations, a pronounced peak at the N\'{e}el ordering vector $\left(\pi,\pi\right)$ appears if both $J_2/J_1$ and $J_3/J_1$ are small, while at large $J_2/J_1$ but small $J_3/J_1$ the structure factor is peaked at the columnar ordering vector $\left(\pi,0\right)$. For large $J_3/J_1$, finally, the peak is located at $\left(q,q\right)$, where $q$ tends to $\pi/2$. 

    \begin{figure}
        \centering
        \includegraphics[width=0.49\textwidth]{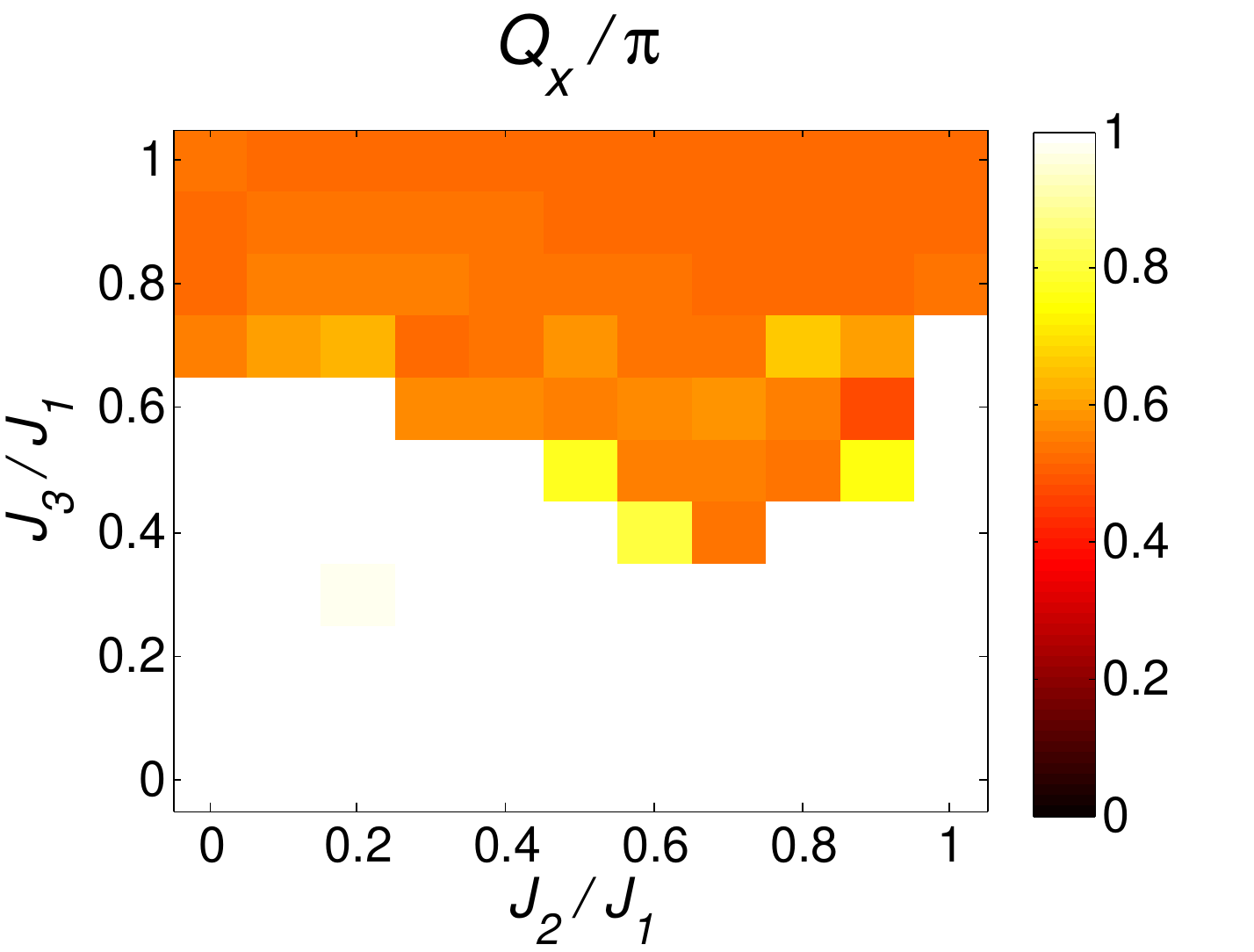} \includegraphics[width=0.49\textwidth]{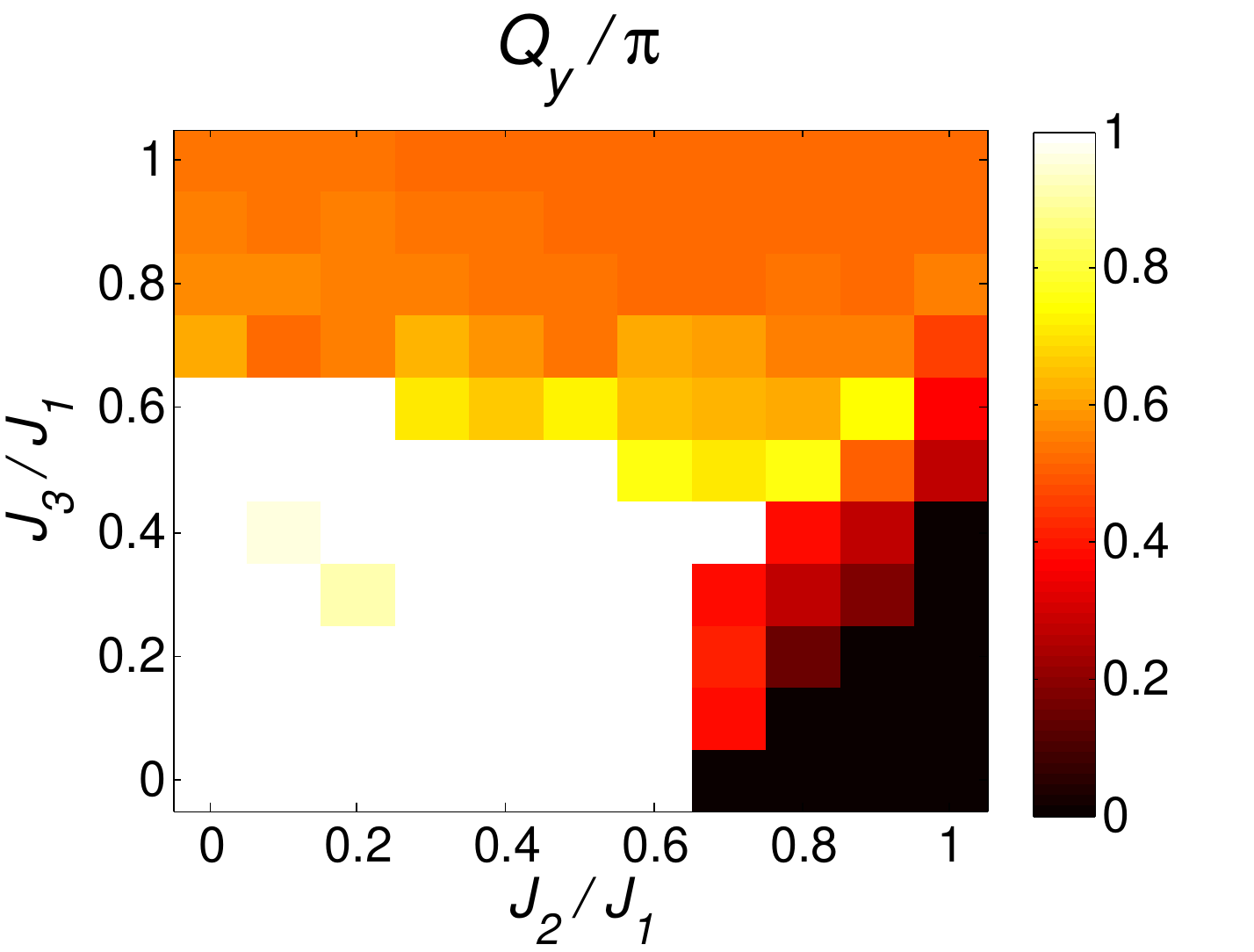}%\\
        \caption{
        Components of the ordering vector for a PEPS calculation on a $8\times 8$ lattice with auxiliary dimension $D=3$.}
        \label{fig:J1J2J3PEPS_Q}
    \end{figure}

%%%%%%%%%%%%%%%%%%%%%%%%%%%%%%%%%%%%%%%%%%%%%%%

\section{Conclusion 		\label{cha:conclusion}}

In this work, we made use of Takahashi's modified spin-wave theory with ordering vector optimization to determine the ground state phase diagram of 
two paradigmatic models of two-dimensional frustrated antiferromagnetism:
the $S=1/2$ Heisenberg model on the SATL and on the $J_1 J_2 J_3$ lattice.
The optimization of the ordering vector shows dramatic quantum corrections to the ordering vector for spiraling states present in both models: 
such corrections show the general trend of promoting collinearly ordered states (either N\'eel or columnar states) against
spiraling ones.    
Both for the triangular and the $J_1 J_2 J_3$ lattice, MSW theory breaks down over a sizable region of parameter space,
showing a dramatic suppression of the order parameter and of the spin stiffness as the breakdown region 
is approached: this finding is strongly suggestive of the appearance of quantum-disordered regions in 
the phase diagram of the models under investigation, an issue which is still under intense debate. 
The extent of the quantum-disordered regions estimated via MSW theory generally appears to be 
lower than that estimated by more accurate numerical techniques which take into account quantum 
fluctuations in a more complete fashion. Hence, one can draw two main conclusions from our results:  
on the one hand MSW might still converge to a magnetically ordered ground state even though the true 
ground state is disordered -- although in this case it will probably feature a small value for the order
parameter, or a small stiffness, suggesting that the magnetic order is not robust when dealing with 
quantum fluctuations more accurately; on the other hand, the breakdown of MSW theory seems to be 
a strong indication that the true ground state is disordered. 
 
  In particular, in the case of the SATL, MSW theory 
 completely breaks down for sufficiently weak couplings between the chains
 composing the lattice, suggesting that the system remains in a disordered
 1D-like state even when the chains are coupled, as already predicted by 
 recent variational approaches. A further disordered phase appears when the 
 inter-chain couplings exceed the intra-chain ones: this phase is
 sandwiched in between the spiral phase of the nearly isotropic triangular lattice
 and the N\'eel phase appearing at large interchain couplings. 
In the case of the $J_1 J_2 J_3$ lattice, a large breakdown region separates
the N\'eel-ordered region for small $J_2$ and $J_3$, from the columnar-ordered
region for $J_2> J_1/2$ and small $J_3$, and from the spiral phase at large 
$J_3$. Hence, a general conclusion that we can draw from the study of these
two models is that collinearly ordered phases (N\'eel and columnar) 
and spiral phases cannot be connected adiabatically -- at least at the 
MSW level -- but they are always separated by a breakdown region;
this is a signal that in the true ground state collinear and spiral phases
might always be divided by an intermediate quantum-disordered phase.

 Quantitative comparisons with more accurate methods (exact diagonalization, and 
 variational \emph{Ansatzes} based on projected BCS states and projected 
 entangled-pair states) reveal that MSW theory with ordering wave-vector
 optimization goes well beyond
 linear spin-wave theory in dealing with quantum effects, and it 
 correctly accounts for the quantum correction to the ordering wave-vector
 of the ordered phases, and for the strong suppression (or total cancellation)
 of magnetic order in correspondence with the candidate regions for 
 quantum-disordered behavior. Given its flexibility and its modest 
 numerical cost, MSW theory serves therefore as a unique tool for 
 the identification of novel quantum phases in strongly frustrated quantum 
 Heisenberg antiferromagnets.

%%%%%%%%%%%%%%%%%%%%%%%%%%%%%%%%%%%%%%%%%%%%%%%%%%%%%%%%%%%%%%%%%%%%%%%%%%%%%%%%%%%%%%%%%%%%%%%%%%%%%%%%%%%%%%%%%%%%%%%%

\section{Acknowledgments}
Two of us (P. H. and T. R.) acknowledge the hospitality of the Kavli
Institute for Theoretical Physics, where this work was finalized.
This work is financially supported by the Caixa Manresa, Spanish MICINN (FIS2008-00784 and Consolider QOIT), EU Integrated Project AQUTE, the EU STREP NAMEQUAM, and ERC Advanced Grant QUAGATUA.

%%%%%%%%%%%%%%%%%%%%%%%%%%%%%%%%%%%%%%%%%%%%%%%%%%%%%%%%%%%%%%%%%%%%%%%%%%%%%%%%%%%%%%%%%%%%%%%%%%%%%%%%%%%%%%%%%%%%%%%%

\appendix

\section{Modified spin-wave formalism for Heisenberg antiferromagnets\label{app:MSW}}

In this Appendix, we shortly review the MSW formalism as applied to Heisenberg
antiferromagnets. The full description of the approach -- as applied to XY models -- 
can be found in \cite{Hauke2010}.
 
The Dyson--Maleev transformation \cite{Dyson1956,Maleev1957} maps the Heisenberg Hamiltonian, Eq.~\eqref{HS}, to the non-linear bosonic Hamiltonian
\begin{eqnarray}
\label{H4}
\null\hspace*{-2cm}
    {\cal H}&=&\frac 1 4 \sum_{\braket{i,j}} J_{ij} \left\lbrace \phantom{+4} \left[ 2 S \left( a_i^\dagger a_j + a_i a_j^\dagger \right) - a_i^\dagger a_j^\dagger a_j a_j - a_i^\dagger a_i a_i a_j^\dagger \right] \left(1+\cos\left(\vect{Q}\cdot\vect{r}_{ij}\right)\right) \right. \nonumber \\
 \null\hspace*{-2cm}   & &\phantom{\frac 1 4 \sum_{\braket{i,j}} t_{ij} \lbrace 4} + \left[ 2 S \left( a_i^\dagger a_j^\dagger + a_i a_j \right) - a_i a_j^\dagger a_j a_j - a_i^\dagger a_i a_i a_j \right] \left(1-\cos\left(\vect{Q}\cdot\vect{r}_{ij}\right)\right) \\
 \null\hspace*{-2cm}   & &\phantom{\frac 1 4 \sum_{\braket{i,j}} t_{ij} \lbrace 4} \left.+\,4 \left[ S^2 - S\left(a_i^\dagger a_i + a_j^\dagger a_j\right)+a_i^\dagger a_i a_j^\dagger a_j\right] \cos\left(\vect{Q}\cdot\vect{r}_{ij}\right) + {\cal O}\left( \frac{1}{S}\right) \quad \right\rbrace \,, \nonumber
\end{eqnarray}
where $a_i$ ($a_i^{\dagger}$) destroys (creates) a Dyson--Maleev boson at site $i$, $S$ is the length of the spin, and $\vect{Q}$ the ordering vector.
Here, we neglected the kinematic constraint which restricts the Dyson--Maleev-boson density $n$ to the physical subspace $n < 2S$, given by the length of the spins $S$. Moreover, we dropped terms with six boson operators, which are of order  ${\cal O}[n/(2S)^3]$ and are negligible for $n/(2S)<1$.
Using Wick's theorem \cite{Fetter1971}, and defining the correlators $\braket{a_i^\dagger a_j}=F\left( \vect{r}_{ij} \right)-\frac 1 2 \delta_{ij}$ and $\braket{a_i a_j}= \braket{a_i^\dagger a_j^\dagger} \,\, = \,\, G\left( \vect{r}_{ij} \right)$, the expectation value $E\equiv\braket{\cal H}$ can be written as
\begin{eqnarray}
    E=\frac 1 2 \sum_{\braket{i,j}} t_{ij} & & \left\lbrace \left[S+\frac 1 2 - F\left( 0 \right) + F\left( \vect{r}_{ij} \right) \right]^2 \left(1+\cos\left(\vect{Q}\cdot\vect{r}_{ij}\right)\right) \right.\\  \nonumber
    & &\left. - \left[S+\frac 1 2 - F\left( 0 \right) + G\left( \vect{r}_{ij} \right) \right]^2 \left(1-\cos\left(\vect{Q}\cdot\vect{r}_{ij}\right)\right)\,\right\rbrace \,.
\end{eqnarray}
After Fourier transforming, $a_{\vect{k}}=\frac{1}{\sqrt{N}}\sum_i a_i\, \ue^{-i \vect{k}\cdot\vect{r}_i}$, where $N$ is the number of sites, and a subsequent Bogoliubov transformation, $\alpha_{\vect{k}\phantom{-}} = \phantom{-}\cosh\theta_{\vect{k}}\, a_{\vect{k}} - \sinh\theta_{\vect{k}} \, a_{-\vect{k}}^\dagger$, and $\alpha_{-\vect{k}}^\dagger = -\sinh\theta_{\vect{k}} \, a_{\vect{k}} + \cosh\theta_{\vect{k}} \, a_{-\vect{k}}^\dagger$, we 
minimize the free energy under the constraint of vanishing magnetization at each site, $\braket{a_i^{\dagger} a_i} = S$ \cite{Takahashi1989}. (This guarantees that the kinematic constraint is satisfied in the mean.)
This yields a set of self-consistent equations, 
\begin{equation}
    \label{tanh2th}
    \tanh 2\theta_{\vect{k}}=\frac{A_{\vect{k}}}{B_{\vect{k}}}
\end{equation}
with
\begin{subequations}
    \label{AkBk}
\begin{eqnarray}
    \label{Ak}
    A_{\vect{k}} & = & \frac 1 N \sum_{\braket{i,j}} t_{ij} \left(1 - \cos\left({\vect{Q}\cdot\vect{r}_{ij}}\right)\right) G_{ij} \,\ue^{i\vect{k}\cdot\vect{r}_{ij}}\,,\\
    \label{Bk}
    B_{\vect{k}} &= &\frac 1 N \sum_{\braket{i,j}} t_{ij} \left[ \left(1 - \cos\left({\vect{Q}\cdot\vect{r}_{ij}}\right)\right) G_{ij} \right. \\
    & & \left. - \left(1 + \cos\left({\vect{Q}\cdot\vect{r}_{ij}}\right)\right) F_{ij} \left(1-\ue^{i\vect{k}\cdot\vect{r}_{ij}}\right)\right] -\mu \nonumber \,,
\end{eqnarray}
\end{subequations}
where $\mu$ is the Lagrange multiplier for the constraint.
The spin-wave spectrum reads
\begin{equation}
\label{disp}
\omega_{\vect{k}}=\sqrt{B_{\vect{k}}^2-A_{\vect{k}}^2}\,.
\end{equation}
At $T=0$, where $n_{\vect{k}}=0\,\, \forall \vect{k}\neq 0$, one finds that $\mu$ vanishes. This implies also the disappearance of the gap at $\vect{k}=0$ that may exist for finite temperature. A vanishing gap is a necessary condition for magnetic LRO.
It also enables Bose condensation in the $\vect{k}=0$ mode. 
Separating out the contribution of the zero mode, 
$\braket{a_{\vect{k}=0}^{\dagger} a_{\vect{k}=0}}/N=\braket{a_{\vect{k}=0} a_{\vect{k}=0}}/N\equiv M_0$
(corresponding to the magnetic order parameter), one arrives at the zero-temperature equations
\begin{subequations}
\label{FG}
\begin{eqnarray}
    F_{ij}&=&M_0 + \frac 1 {2 N} \sum_{\vect{k}\neq 0} \frac{B_{\vect{k}}} {\omega_{\vect{k}}}\cos\left(\vect{k}\cdot\vect{r}_{ij}\right)\label{Fij}\,,\\
    G_{ij}&=&M_0 + \frac 1 {2 N} \sum_{\vect{k}\neq 0} \frac{A_{\vect{k}}} {\omega_{\vect{k}}}\cos\left(\vect{k}\cdot\vect{r}_{ij}\right)\label{Gij}\,,
\end{eqnarray}
\end{subequations}
and the constraint of vanishing magnetization at each site becomes
\begin{equation}
    \label{constr2}
    S+\frac 1 2 = M_0 + \frac 1 {2 N} \sum_{\vect{k}\neq 0} \frac{B_{\vect{k}}} {\omega_{\vect{k}}}.
\end{equation}

It is not \emph{a priori} clear that the classical ordering vector $\vect{Q}^{\mathrm{cl}}$ correctly describes the LRO in the quantum system. 
To account for the competition between states with LRO at different ordering vectors $\vect{Q}$ we extend the MSW procedure by optimizing the free energy $\mathcal{F}$ with respect to the ordering vector $\vect{Q}$.
This yields two additional equations which must be added to the set of self-consistent equations, 
\begin{subequations}
\label{Qs}
\begin{equation}
    \label{Qx}
    \frac\partial{\partial Q_x} \mathcal{F}=-\frac 1 2 \sum_{\braket{i,j}} t_{ij} \sin\left(\vect{Q}\cdot\vect{r}_{ij}\right)r_{ij}^x\left[F_{ij}^2+G_{ij}^2\right] = 0 \,,
\end{equation}
\begin{equation}
    \label{Qy}
    \frac\partial{\partial Q_y} \mathcal{F}=-\frac 1 2 \sum_{\braket{i,j}} t_{ij} \sin\left(\vect{Q}\cdot\vect{r}_{ij}\right)r_{ij}^y\left[F_{ij}^2+G_{ij}^2\right]=0\,.
\end{equation}
\end{subequations}
In the SATL with NN interactions these simplify to
$Q_y=0$ and
\begin{equation}
    \label{Qx2}
    Q_x=2\arccos\left[-\frac \alpha 2 \frac{F_{\vs{\tau}_2}^2+G_{\vs{\tau}_2}^2}{F_{\vs{\tau}_1}^2+G_{\vs{\tau}_1}^2}\right]\,,
\end{equation}
where $\vs{\tau}_1=\left(1,0\right)$ and $\vs{\tau}_2=\left(1/2,\sqrt{3}/2\right)$ are the lattice vectors.

The values of $F_{ij}$ and $G_{ij}$ can now be calculated by solving self-consistently Eqs.~(\ref{AkBk}--\ref{Qs}).
Through Wick's theorem the knowledge of the quantities $F_{ij}$ and $G_{ij}$ 
allows the computation of the expectation value of any observable.

\paragraph{Spin stiffness\label{cha:gausscurvtriang}}

The optimization of the ordering vector allows a straightforward calculation of the spin stiffness, which gives a measure of how stiff magnetic LRO order is with respect to distortions of the ordering vector, and thus provides a fundamental self-consistency check of our approach. In fact, finding a small spin stiffness casts doubt on the reliability of the spin-wave approach in describing such a strongly fluctuating state, and hence suggests that the true ground state might be quantum disordered. 

The spin stiffness tensor is defined as $\rho_{\alpha\beta}=\frac 1 N \left. \frac{\ud^2 \mathcal{F}}{\ud Q_\alpha \ud Q_\beta}\right|_{\vect{Q}=\vect{Q^0}}$, 
evaluated at the optimized ordering vector $\vect{Q^0}$. From this we can extract the \emph{parallel spin stiffness}
$
\rho_{\|}\equiv\frac{1}{2}\left(\rho_{xx}+\rho_{yy}\right)
$
and the \emph{Gaussian spin stiffness} $\Upsilon=\det \rho$.

Since a change in $\vect{Q}$ affects the correlators $F_{ij}$ and $G_{ij}$, we must compute $\Upsilon$ self-consistently. After finding the optimal $\vect{Q}^0$ by the self-consistent procedure described above, we calculate $\frac{1}{N} \mathcal{F}\left(Q_x,Q_y\right)$ self-consistently for several fixed ordering vectors $\vect{Q}=\vect{Q}^0+\Delta\vect{Q}$ and fit a quadratic form to the results. Since the minimum in the free energy can be very shallow, this procedure can be affected by numerical noise. 
As an approximation to the true spin stiffness, the \emph{partial spin stiffness} $\rho_{\alpha\beta}^{\mathrm{partial}}$ can be computed via the partial derivatives, \emph{i.e.}, without recalculating the self-consistent equations. It reads
\begin{eqnarray}
    \label{gcpartial}
    \rho_{\alpha\beta}^{\mathrm{partial}}&\equiv&\frac 1 N \frac{\partial^2}{\partial Q_{\alpha}\partial Q_{\beta}} \mathcal{F}\\
    &=&-\frac 1 {2N} \sum_{\braket{i,j}} t_{ij} \cos\left(\vect{Q}\cdot\vect{r}_{ij}\right)r_{ij}^{\alpha}r_{ij}^{\beta}\left[F_{ij}^2+G_{ij}^2\right] \nonumber \,.
\end{eqnarray}
We define $\Upsilon^{\mathrm{partial}}$ analogously to $\Upsilon$ as the determinant of the partial spin-stiffness tensor.\\

\bibliographystyle{nature}
\bibliography{D:/work/Diplom/thesis/references}

\begin{thebibliography}{10}

\bibitem{Dyson1978}
Dyson, F.~J., Lieb, E.~H., and Simon, B.
\newblock {\em J. Stat. Phys.}{ \bf 18}, 335 (1978).

\bibitem{Kennedy1988}
Kennedy, T., Lieb, E.~H., and Shastry, B.~S.
\newblock {\em J. Stat. Phys.}{ \bf 53}, 1019 (1988).

\bibitem{Manousakis1991}
Manousakis, E.
\newblock {\em Rev. Mod. Phys.}{ \bf 63}, 1 (1991).

\bibitem{Misguich2004}
Misguich, G. and Lhuillier, C.
\newblock {\em Frustrated Spin Systems},  229.
\newblock World Scientific, Singapore (2004).

\bibitem{Anderson1973}
Anderson, P.
\newblock {\em Materials Research Bulletin}{ \bf 8}, 153 (1973).

\bibitem{Fazekas1974}
Fazekas, P. and Anderson, P.~W.
\newblock {\em Philosophical Magazine}{ \bf 30}, 423 (1974).

\bibitem{Kastner1998}
Kastner, M.~A., Birgeneau, R.~J., Shirane, G., and Endoh, Y.
\newblock {\em Rev. Mod. Phys.}{ \bf 70}, 897 (1998).

\bibitem{Lee2006}
Lee, P.~A., Nagaosa, N., and Wen, X.-G.
\newblock {\em Rev. Mod. Phys.}{ \bf 78}, 17 (2006).

\bibitem{delaCruz2008}
de~la Cruz, C., Huang, Q., Lynn, J.~W., Li, J., {Ratcliff II}, W., Zarestky,
  J.~L., Mook, H.~A., Chen, G.~F., Luo, J.~L., Wang, N.~L., and Dai, P.
\newblock {\em Nature}{ \bf 453}, 899 (2008).

\bibitem{Coldea2001}
Coldea, R., Tennant, D.~A., Tsvelik, A.~M., and Tylczynski, T.
\newblock {\em Phys. Rev. Lett.}{ \bf 86}, 1335 (2001).

\bibitem{Shimizu2003}
Shimizu, Y., Miyagawa, K., Kanoda, K., Maesato, M., and Saito, G.
\newblock {\em Phys. Rev. Lett.}{ \bf 91}, 107001 (2003).

\bibitem{Yamashita2008}
Yamashita, S., Nakazawa, Y., Oguni, M., Oshima, Y., Nojiri, H., Shimizu, Y.,
  Miyagawa, K., and Kanoda, K.
\newblock {\em Nat. Phys.}{ \bf 4}, 459 (2008).

\bibitem{Carretta2004}
Carretta, P., Papinutto, N., Melzi, R., Millet, P., Gonthier, S., Mendels, P.,
  and Wzietek, P.
\newblock {\em J. Phys. Condens. Matter}{ \bf 16}, S849 (2004).

\bibitem{Nath2008}
Nath, R., Tsirlin, A.~A., Rosner, H., and Geibel, C.
\newblock {\em Phys. Rev. B}{ \bf 78}, 064422 (2008).

\bibitem{Weihong1999}
Weihong, Z., McKenzie, R.~H., and Singh, R. R.~P.
\newblock {\em Phys. Rev. B}{ \bf 59}, 14367 (1999).

\bibitem{Yunoki2006}
Yunoki, S. and Sorella, S.
\newblock {\em Phys. Rev. B}{ \bf 74}, 014408 (2006).

\bibitem{Weng2006}
Weng, M.~Q., Sheng, D.~N., Weng, Z.~Y., and Bursil, R.~J.
\newblock {\em Phys. Rev. B}{ \bf 74}, 012407 (2006).

\bibitem{Fjaerestad2007}
Fjaerestad, J.~O., Zheng, W., Singh, R. R.~P., McKenzie, R.~H., and Coldea, R.
\newblock {\em Phys. Rev. B}{ \bf 75}, 174447 (2007).

\bibitem{Kohno2007}
Kohno, M., Starykh, O.~A., and Balents, L.
\newblock {\em Nat. Phys.}{ \bf 3}, 790 (2007).

\bibitem{Starykh2007}
Starykh, O.~A. and Balents, L.
\newblock {\em Phys. Rev. Lett.}{ \bf 98}, 077205 (2007).

\bibitem{Heidarian2009}
Heidarian, D., Sorella, S., and Becca, F.
\newblock {\em Phys. Rev. B}{ \bf 80}, 012404 (2009).

\bibitem{Singh1999}
Singh, R. R.~P., Weihong, Z., Hamer, C.~J., and Oitmaa, J.
\newblock {\em Phys. Rev. B}{ \bf 60}, 7278 (1999).

\bibitem{Capriotti2001}
Capriotti, L., Becca, F., Parola, A., and Sorella, S.
\newblock {\em Phys. Rev. Lett.}{ \bf 87}, 097201 (2001).

\bibitem{Sushkov2001}
Sushkov, O.~P., Oitmaa, J., and Weihong, Z.
\newblock {\em Phys. Rev. B}{ \bf 63}, 104420 (2001).

\bibitem{Sindzingre2004}
Sindzingre, P.
\newblock {\em Phys. Rev. B}{ \bf 69}, 094418 (2004).

\bibitem{Sirker2006}
Sirker, J., Weihong, Z., Sushkov, O.~P., and Oitmaa, J.
\newblock {\em Phys. Rev. B}{ \bf 73}, 184420 (2006).

\bibitem{Mambrini2006}
Mambrini, M., L{\"a}uchli, A., Poilblanc, D., and Mila, F.
\newblock {\em Phys. Rev. B}{ \bf 74}, 144422 (2006).

\bibitem{Darradi2008}
Darradi, R., Derzhko, O., Zinke, R., Schulenburg, J., Krueger, S.~E., and
  Richter, J.
\newblock {\em Phys. Rev. B}{ \bf 78}, 214415 (2008).

\bibitem{Takahashi1989}
Takahashi, M.
\newblock {\em Phys. Rev. B}{ \bf 40}, 2494 (1989).

\bibitem{Xu1991}
Xu, J.~H. and Ting, C.~S.
\newblock {\em Phys. Rev. B}{ \bf 43}, 6177 (1991).

\bibitem{Hauke2010}
Hauke, P., Roscilde, T., Murg, V., Cirac, J.~I., and Schmied, R.
\newblock {\em New J. Phys.}{ \bf 12}, 053036 (2010).

\bibitem{Figueirido1989}
Figueirido, F., Karlhede, A., Kivelson, S., Sondhi, S., Rocek, M., and Rokhsar,
  D.~S.
\newblock {\em Phys. Rev. B}{ \bf 41}, 4619 (1989).

\bibitem{Read1991}
Read, N. and Sachdev, S.
\newblock {\em Phys. Rev. Lett.}{ \bf 66}, 1773 (1991).

\bibitem{Ferrer1993}
Ferrer, J.
\newblock {\em Phys. Rev. B}{ \bf 47}, 8769 (1993).

\bibitem{Manuel1999}
Manuel, L.~O. and Ceccatto, H.~A.
\newblock {\em Phys. Rev. B}{ \bf 60}, 9489 (1999).

\bibitem{Schmied2008}
Schmied, R., Roscilde, T., Murg, V., Porras, D., and Cirac, J.~I.
\newblock {\em New J. Phys.}{ \bf 10}, 045017 (2008).

\bibitem{Richter2010}
Schulenburg, J. and Richter, J.
\newblock {\em Eur. Phys. J. B}{ \bf 73}, 117 (2010).

\bibitem{Weber2006}
Weber, C., L\"{a}uchli, A., Mila, F., and Giamarchi, T.
\newblock {\em Phys. Rev. B}{ \bf 73}, 014519 (2006).

\bibitem{Singh1989a}
Singh, R. R.~P.
\newblock {\em Phys. Rev. B}{ \bf 39}, 9760 (1989).

\bibitem{Capriotti1999a}
Capriotti, L., Trumper, A.~E., and Sorella, S.
\newblock {\em Phys. Rev. Lett.}{ \bf 82}, 3899 (1999).

\bibitem{Sandvik1997}
Sandvik, A.
\newblock {\em Phys. Rev. B}{ \bf 56}, 11678 (1997).

\bibitem{Trumper1999}
Trumper, A.~E.
\newblock {\em Phys. Rev. B}{ \bf 60}, 2987 (1999).

\bibitem{Lecheminant1995}
Lecheminant, P., Bernu, B., Lhuillier, C., and Pierre, L.
\newblock {\em Phys. Rev. B}{ \bf 52}, 9162 (1995).

\bibitem{Shastry1990}
Shastry, B.~S. and Sutherland, B.
\newblock {\em Phys. Rev. Lett.}{ \bf 65}, 243 (1990).

\bibitem{Kawamura2002}
Kawamura, H.
\newblock {\em arXiv:cond-mat/0202109v1}{ \bf } (2002).

\bibitem{Richter1991}
Richter, J., Gros, C., and Weber, W.
\newblock {\em Phys. Rev. B}{ \bf 44}, 906 (1991).

\bibitem{Gelfand1989}
Gelfand, M.~P., Singh, R.~R., and Huse, D.~A.
\newblock {\em Phys. Rev. B}{ \bf 40}, 10801 (1989).

\bibitem{Moreo1990}
Moreo, A., Dagotto, E., Jolicoeur, T., and Riera, J.
\newblock {\em Phys. Rev. B}{ \bf 42}, 6283 (1990).

\bibitem{Chubukov1991}
Chubukov, A.
\newblock {\em Phys. Rev. B}{ \bf 44}, 392 (1991).

\bibitem{Leung1996}
Leung, P.~W. and Lam, N.
\newblock {\em Phys. Rev. B}{ \bf 53}, 2213 (1996).

\bibitem{Chandra1988}
Chandra, P. and Doucot, B.
\newblock {\em Phys. Rev. B}{ \bf 38}, 9335 (1988).

\bibitem{Locher1990}
Locher, P.
\newblock {\em Phys. Rev. B}{ \bf 41}, 2537 (1990).

\bibitem{Zhong1993}
Zhong, Q.~F. and Sorella, S.
\newblock {\em Europhys. Lett.}{ \bf 21}, 629 (1993).

\bibitem{Capriotti2004a}
Capriotti, L., Scalapino, D.~J., and White, S.~R.
\newblock {\em Phys. Rev. Lett.}{ \bf 93}, 177004 (2004).

\bibitem{Capriotti2004b}
Capriotti, L. and Sachdev, S.
\newblock {\em Phys. Rev. Lett.}{ \bf 93}, 257206 (2004).

\bibitem{Murg2009}
Murg, V., Verstraete, F., and Cirac, J.~I.
\newblock {\em Phys. Rev. B}{ \bf 79}, 195119 (2009).

\bibitem{Barabanov1990}
Barabanov, A.~F. and Starykh, O.~A.
\newblock {\em JETP Lett.}{ \bf 51}, 312 (1990).

\bibitem{Xu1990}
Xu, J.~H. and Ting, C.~S.
\newblock {\em Phys. Rev. B}{ \bf 42}, 6861 (1990).

\bibitem{Ivanov1992}
Ivanov, N.~B. and Ivanov, P.~C.
\newblock {\em Phys. Rev. B}{ \bf 46}, 8206 (1992).

\bibitem{Gochev1994}
Gochev, I.~G.
\newblock {\em Phys. Rev. B}{ \bf 49}, 9594 (1994).

\bibitem{Dotsenko1994}
Dotsenko, A.~V. and Sushkov, O.~P.
\newblock {\em Phys. Rev. B}{ \bf 50}, 13821 (1994).

\bibitem{Einarsson1995}
Einarsson, T. and Schulz, H.~J.
\newblock {\em Phys. Rev. B}{ \bf 51}, 6151 (1995).

\bibitem{Schulz1996}
Schulz, H., Ziman, T., and Poilblanc, D.
\newblock {\em J. Phys. I France}{ \bf 6}, 675 (1996).

\bibitem{Trumper1997}
Trumper, A.~E., Manuel, L.~O., Gazza, C.~J., and Ceccatto, H.~A.
\newblock {\em Phys. Rev. Lett.}{ \bf 78}, 2216 (1997).

\bibitem{Manuel1998}
Manuel, L.~O., Trumper, A.~E., and Ceccatto, H.~A.
\newblock {\em Phys. Rev. B}{ \bf 57}, 8348 (1998).

\bibitem{Murg2008}
Murg, V., Verstraete, F., and Cirac, J.
\newblock {\em to be published}{ \bf } (2009).

\bibitem{Bonca1994}
Bon\v{c}a, J., Rodriguez, J.~P., Ferrer, J., and Bedell, K.~S.
\newblock {\em Phys. Rev. B}{ \bf 50}, 3415 (1994).

\bibitem{Dyson1956}
Dyson, F.~J.
\newblock {\em Phys. Rev.}{ \bf 102}, 1217 (1956).

\bibitem{Maleev1957}
Maleev, S.~V.
\newblock {\em Zh. Eksp. Teor. Fiz.}{ \bf 30}, 1010 (1957).
\newblock see also Sov. Phys. JETP 6, 776 (1958).

\bibitem{Fetter1971}
Fetter, A. and Walecka, J.
\newblock {\em Quantum Theory of Many-Particle Systems}.
\newblock McGraw Hill, New York,  (1971).

\end{thebibliography}
%\bibliography{references}

\end{document}